\label{Options}
\documentclass[epj]{svjour}

\label{Packages}
\usepackage{graphicx}
\usepackage{color}
\usepackage{amsmath}
\usepackage{amssymb}

\begin{document}
\label{Commands}
\newcommand{\be}{\begin{equation}}
\newcommand{\ee}{\end{equation}}
\newcommand{\highlight}[1]{\colorbox{yellow}{#1}}
\newcommand{\bS}{{\bf S}}
\newcommand{\mP}{{\mathcal P}}
\newcommand{\mT}{{\mathcal T}}
\newcommand{\mC}{{\mathcal C}}
\newcommand{\bd}{{\bf d}}
\newcommand{\bB}{{\bf B}}
\newcommand{\bQ}{{\bf Q}}
\newcommand{\bk}{{\bf k}}
\newcommand{\bK}{{\bf K}}
\newcommand{\bv}{{\bf v}}
\newcommand{\bl}{{\bf l}}
\newcommand{\br}{{\bf r}}
\newcommand{\bp}{{\bf p}}
\newcommand{\bq}{{\bf q}}
\newcommand{\s}{{\sigma}}
\newcommand{\bA}{{\bf A}}
\newcommand{\ba}{{\bf a}}
\newcommand{\bJ}{{\bf J}}
\newcommand{\bE}{{\bf E}}
\newcommand{\bb}{{\bf b}}
\newcommand{\bsig}{\mbox{\boldmath{$\sigma$}}}
\newcommand{\Det}{\,{\rm{Det}\,}}
\newcommand{\Tr}{\,{\rm{Tr}\,}}
\newcommand{\tr}{\,{\rm{tr}\,}}
\newcommand{\bn}{{\bf \nabla}}
\newcommand{\sign}{\,{\rm{sign}\,}}
\newcommand{\bomega}{\bar{\Omega}}
\newcommand{\wL}{\omega_{L}}
\newcommand{\COMMENT}{\large\bf}
\newcommand{\FigWidth}{\columnwidth}
\newcommand{\beqa}{\begin{eqnarray}}
\newcommand{\eeqa}{\end{eqnarray}}
\newcommand{\bI}{\bf{I}_s}
\renewcommand{\Re}{{\rm Re}}
\renewcommand{\Im}{{\rm Im}}

%**************************************************************************************

\title{Optical waveguide arrays: quantum effects and PT symmetry breaking}
\author{Yogesh N. Joglekar\inst{1}, Clinton Thompson\inst{1}, Derek D. Scott\inst{1}, Gautam Vemuri\inst{1}}
\institute{Department of Physics, Indiana University - Purdue University Indianapolis (IUPUI), 
Indianapolis, Indiana 46202, USA}
\date{Received: date / Revised version: date}
\abstract{
Over the last two decades, advances in fabrication have led to significant progress in creating patterned heterostructures that support either carriers, such as electrons or holes, with specific band structure or electromagnetic waves with a given mode structure and dispersion. In this article, we review the properties of light in coupled optical waveguides that support specific energy spectra, with or without the effects of disorder, that are well-described by a Hermitian tight-binding model. We show that with a judicious choice of the initial wave packet, this system displays the characteristics of a quantum particle, including transverse photonic transport  and localization, and that of a classical particle. We extend the analysis to non-Hermitian, parity and time-reversal ($\mathcal{PT}$) symmetric Hamiltonians which physically represent waveguide arrays with spatially separated, balanced absorption or amplification. We show that coupled waveguides are an ideal candidate to simulate $\mathcal{PT}$-symmetric Hamiltonians and the transition from a purely real energy spectrum to a spectrum with complex conjugate eigenvalues that occurs in them.} %end of abstract

\titlerunning{Optical waveguide arrays}
\authorrunning{Y.N. Joglekar et al.}
\maketitle

%**************************************************************************************
%**************************************************************************************

\section{Introduction}
\label{intro}
Historically, light and matter have been considered two quintessentially different entities. Since the advent of quantum theory, which elucidates the wave nature of material particles and the particle nature of electromagnetic waves, properties of quantum system of particles are described by a (possibly many-body) wave function whose time evolution is determined by the Schr\"{o}dinger equation~\cite{qm1,qm2}. Such many-body condensed matter systems support collective excitations whose energy is linearly proportional to the momentum, and thus allow one to mimic light - linearly dispersing massless excitations - in material systems~\cite{wen}. However, due to the unique nature of electromagnetic waves, namely the lack of a rest-frame or, equivalently, zero rest mass, they were not considered useful for simulating the behavior of quantum particles with nonzero rest mass~\cite{jackson}. 

Over the past decade, the tremendous progress in fabrication and characterization of semiconductor heterostructures has made it possible to create arrays of evanescently coupled optical waveguides with numbers varying from a few  to a few hundred~\cite{naturereview,discreteo}. The resulting ``diffraction management''~\cite{dm} makes evanescently coupled waveguides a paradigm for the realization of a quantum particle hopping on one or two dimensional lattices, and permits the observation of quantum and condensed matter phenomena in macroscopic samples using electromagnetic waves. One can engineer such a waveguide array to model any desired form of tight-binding, non-interacting Hamiltonian, because the local index of refraction and the width of the waveguide determine the on-site potential for the Hamiltonian while the tunneling amplitude from one site to its adjacent site can be changed by changing the separation between adjacent waveguides~\cite{Jones1965,yariv,review}. A variation in the index of refraction or the tunneling amplitude, both of which can be introduced easily, permit the modeling of a tight-binding Hamiltonian with site or bond disorders respectively. Due to this versatility, many quantum and condensed matter phenomena - Bloch oscillations \cite{Peschel1998},  Dirac zitterbewegung \cite{Longhi2010}, and increased intensity fluctuations~\cite{Lahini2008,Thompson2010} of light undergoing Anderson localization~\cite{all1,all2,all3} - have been theoretically predicted to occur or experimentally observed in waveguide arrays. They have been used to investigate solitonic solutions that arise due to nonlinearities in the dielectric response~\cite{soliton1,soliton2}. Such arrays of coupled waveguides have also been used to simulate the quantum walks of a single photon \cite{Perets2008,qw}, correlated photons \cite{Peruzzo2010}, and Hanbury Brown and Twiss (HBT) correlations~\cite{hbt2010}. Most recently, they have been used to create a ``topological insulator'', an exotic state of matter in which the bulk is an insulator, but the two surfaces are conductors~\cite{ti1,ti2}. 

There are several advantages to using waveguides to investigate quantum behavior and statistics. First, the quantum effects are measurable over much longer distances than those in condensed matter systems with electrons or in cold-atom systems in electromagnetic traps. Second, instead of an indirect measurement through observables such as conductivity or other response functions~\cite{ventra}, in optical waveguides, one can directly measure the time-evolution of a wave function via the time-and-space dependent probability distribution, since it is identical to the light intensity distribution. For lattice models realized via electronic or cold-atom systems, typically, eigenstates in a small fraction of the energy band near the Fermi energy are experimentally probed~\cite{coldatoms}; in contrast, the ability to create an initial wave packet localized to a single waveguide - by coupling light into a single waveguide - means that quantum effects across the entire energy band of the tight-binding model can be investigated in optical waveguide arrays. 

In the past fifteen years, there has been significant theoretical research on properties of non-Hermitian Hamiltonians that, sometimes, show purely real spectra~\cite{Bender1998,Bender2002,ptreview1,ptreview2}. In continuum models, such Hamiltonians usually consist of a Hermitian kinetic energy term and a complex potential that is invariant under the combined operation of parity and time-reversal ($\mathcal{PT}$), such as $V(x)=x^2(ix)^\epsilon$ or $V(x)=n_R(x)+in_I(x)$ where $n_R(x)$ and $n_I(x)$ are even and odd functions of $x$, respectively. The region of parameter space where all energy eigenvalues of a $\mathcal{PT}$-symmetric Hamiltonian are real is traditionally called the $\mathcal{PT}$-symmetric region, and the emergence of complex conjugate eigenvalues that accompanies departure from this region is called $\mathcal{PT}$-symmetry breaking. Since the effective potential in an optical waveguide array is given by the local (complex) index of refraction,  properties of $\mathcal{PT}$ Hamiltonians have led to predictions of new optical phenomenon such as Bloch oscillations in complex crystals~\cite{Longhi2009}, an optical medium that can simultaneously act as an emitter and a perfect absorber of coherent waves~\cite{cabsorb1,cabsorb2}, $\mathcal{PT}$-symmetric Dirac equation~\cite{ptdirac}, and induced quantum coherence between Bose-Einstein condensates~\cite{Xiong2010}. The $\mathcal{PT}$-symmetry breaking has recently been experimentally observed in two coupled waveguides~\cite{Guo2009,Ruter2010}, silicon photonic circuits~\cite{feng}, and optical networks~\cite{ptsynthetic}. Thus, coupled optical waveguide arrays are also an ideal candidate to simulate the quantum dynamics of a non-Hermitian, $\mathcal{PT}$-symmetric Hamiltonian. 

In this paper, we review properties of coupled optical waveguides. In the absence of any loss or gain in a waveguide, the effective Hamiltonian of such an array is Hermitian. In Sec.~\ref{sec:tb} we present the basics of such Hermitian, tight-binding models, and discuss quantum photonic transport (Sec.~\ref{ss:bloch}), continuum quasiclassical limit (Sec.~\ref{ss:cont}), arrays with position-dependent nearest-neighbor tunneling (Sec.~\ref{ss:parity}), and the effects of on-site and tunneling disorder (Sec.~\ref{ss:disorder}). Section~\ref{sec:pt} focuses on $\mathcal{PT}$-symmetric tight-binding models where the non-Hermitian, $\mathcal{PT}$-symmetric potential corresponds to loss in one waveguide and an equal gain in its mirror-symmetric counterpart waveguide. We introduce the terminology, present the $\mathcal{PT}$-symmetric phase diagram for arrays with open boundary conditions (Sec.~\ref{ss:ptphase}), discuss the salient features of non-unitary time evolution in such systems (Sec.~\ref{ss:intensity}), and compare the effects of Hermitian vs. non-Hermitian, $\mathcal{PT}$-symmetric disorder on intensity correlations (Sec.~\ref{ss:ptdisorder}). We conclude this review with a brief discussion of open questions in Sec.~\ref{sec:disc}. 

%**************************************************************************************
%**************************************************************************************

\section{Hermitian Tight-binding Models}
\label{sec:tb}
% Figure: schematic of a waveguide with details.
\begin{figure}[htbp]
\includegraphics[width=\columnwidth]{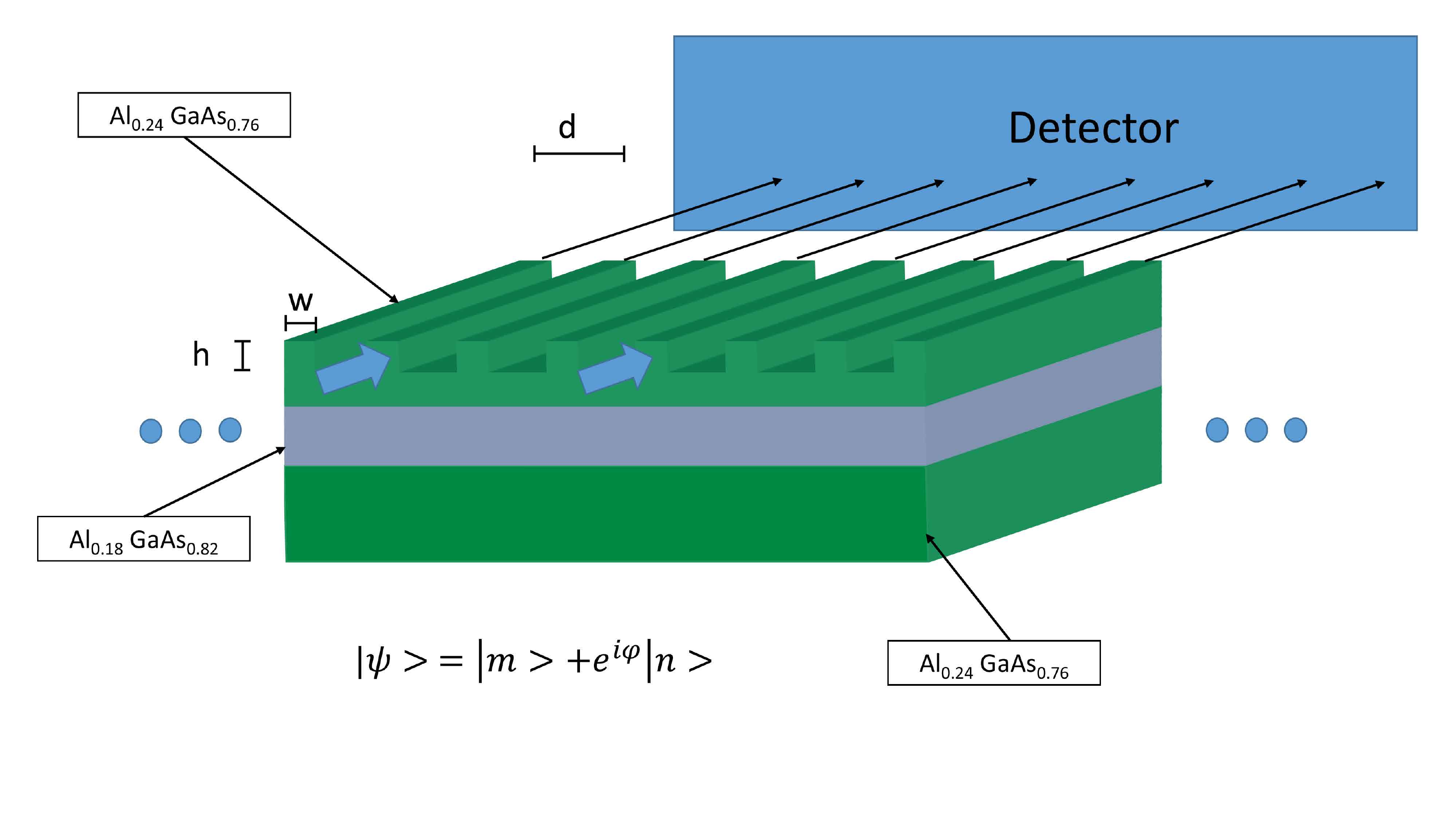}
\caption{Schematic of an array of evanescently coupled optical waveguides. The height $h$ and the width $w$ of the waveguide determine the spatial profile of electromagnetic modes 
inside it, along with the effective potential $\hbar\beta_j$ in waveguide $j$, and the distance $d$ between the centers of adjacent waveguides determines the effective tunneling $\hbar C_{j+1,j}$ between them. Due to its constant speed, the motion of light along the waveguide is equivalent to its time evolution whereas motion across different waveguides simulates a quantum particle on a tight-binding lattice.}
\label{fig:schematic}
\vspace{-5mm}
\end{figure} 

The Hamiltonian for a one-dimensional array with $N$ identical, single-mode waveguides is given by 
\begin{equation}
\label{eq:tbh}
H=\hbar \sum_{j=1}^{N} \left[\beta_{j}a_{j}^{\dag }a_{j}-(C_{j+1,j} a_{j+1}^{\dag }a_{j}+ C_{j,j+1} a_{j}^{\dag }a_{j+1})\right]
\end{equation}
where $\hbar=h/(2\pi)$ is the scaled Planck's constant, $a^\dagger_j (a_j)$ is the Bosonic creation (annihilation) operator for the single mode in waveguide $j$, $\beta_{j}$ is the effective potential on site $j$ or equivalently, the propagation constant for waveguide $j$, and $C_{j+1,j}$ denotes the tunneling amplitude from site $j$ to adjacent site $j+1$. Based upon its geometry, the array can have open boundary conditions ($C_N=0$) or periodic boundary conditions, $a^\dagger_{N+1}=a^\dagger_1$. It is straightforward to generalize this Hamiltonian to two-dimensional arrays. The on-site potential $\beta_j$ and the tunneling amplitude $C_{j+1,j}$ are determined by profile of the electric field $u(\br)$ in a single waveguide as 
\begin{eqnarray}
\label{eq:beta}
\beta_j & = & c\sqrt{k_0^2-\bk_j^2},\\
\label{eq:tunnel}
C_{j+1,j} & = & (n_{j+1}^2-n_b^2)\frac{k_0^2}{2\beta_j}\int_\square d\br\, u_{j+1}(\br) u_j(\br),
\end{eqnarray}
where $k_0$ is the wavenumber for the incident light, $c$ is the speed of light in vacuum, $\bk_j$ characterizes the wave vector for the single eigenmode in waveguide $j$, $n_{j+1}$ and $n_b$ are refractive indices for waveguide $j+1$ and the barrier between adjacent waveguides respectively, and $\square$ denotes integral over the two-dimensional cross section of waveguide $j$. Thus, the potential $\beta_j$ is linearly proportional to the local index of refraction $n_j$, whereas the tunneling amplitude is proportional to the overlap between the electric-field envelope functions in waveguides $j$ and $j+1$. Note that it is possible to create a non-Hermitian tunneling profile - $C_{j+1.j}\neq C_{j,j+1}$ - by varying the index of refraction and maintaining the waveguide geometry; we will, however, only consider waveguide arrays where the tunneling is Hermitian, $C_{j+1,j}=C_{j,j+1}=C_j$. The electromagnetic waves in dielectric media do not interact with each other when the light intensity is small and the effects of non-linear susceptibility $\chi_3$ can be ignored~\cite{nlqo}; therefore, there are no quartic ``interaction terms'' in the Hamiltonian. Thus, the Hamiltonian that describes the time-evolution of an electromagnetic pulse (with many photons) in an array of waveguides is equivalent to that of a {\it single} quantum particle hopping on a lattice with on-site potentials $\beta_j$ and tunneling amplitudes $C_j$. This absence of interaction allows us to use the coupled waveguide array as an exquisite probe of competition among dispersion, disorder, quantum statistics, and boundary conditions.  

When the on-site potential is constant $\beta_j=0$ and tunneling amplitudes are constant, $C_j=C$, the Hamiltonian is translationally invariant. Therefore, it can be diagonalized using eigenfunctions $\psi_{k_n}(j)$ characterized by eigenmomentum $k_n$. The energy spectrum of the one-dimensional lattice is given by $E(k_n)=-\Delta_B\cos(k_n)/2$ where $\Delta_B=4\hbar C$ is the bandwidth and the dimensionless eigenmomenta are $k_n=n\pi/(N+1)$ ($n=1,\ldots,N$) for open boundary conditions and $k_n=\pm 2n\pi/N$ with $n=0,\ldots,N/2$ for periodic boundary conditions~\cite{dse}. It follows then that for an array with $N\rightarrow\infty$ sites and lattice spacing $a$, the permitted dimensionful wave vectors $k$ form a continuum, bounded by $-\pi/a< k\leq \pi/a$, known as the first Brillouin zone~\cite{bz}. 

We emphasize that although electrons in condensed matter materials and light in optical waveguide arrays can both be described by Eq.(\ref{eq:tbh}), the relevant lattice-site numbers and energy scales in the two cases are vastly different. For electronic materials, the number of atoms or lattice sites is $N\gg 10^9$ whereas for light, the number of coupled waveguides is $N\lesssim 100$. For electrons, the tunneling amplitude $\hbar C\sim 1$ eV or equivalently, $C\sim$ 240 THz and $C/(2\pi c)\sim$ 8000 cm$^{-1}$, whereas the on-site potential $\hbar\beta\sim E_F\ll \Delta_B$ where $E_F$ is the Fermi energy; these parameters cannot be varied significantly (by orders of magnitude) since Coulomb interactions are the primary determinant for these parameters. For light, the tunneling amplitude, determined by the distance between adjacent waveguides, is $C/(2\pi c)\sim$ 3-50 cm$^{-1}$. Thus, the typical bandwidth of the waveguide array, $\Delta_B\sim$ a few meV, is smaller than its electronic counterpart by orders of magnitude. In addition, the on-site potential in waveguide arrays can be comparable with the bandwidth, $\hbar\beta\sim\Delta_B\sim$ 10-100 cm$^{-1}$. This tremendous flexibility, present even in a {\it small array with a constant tunneling}, hints at the rich possibilities for designing waveguide arrays with dramatically different properties. In the following subsections, we will illustrate this point with a few examples. 

%**************************************************************************************

\subsection{Phase controlled photonic transport}
\label{ss:bloch}
In free space, the change in the momentum of a particle under constant force, or equivalently, a potential that varies linearly with position, is proportional to the time and thus can increase continuously. In sharp contrast, when a particle on a lattice is acted upon by a constant force, its momentum change is bounded by the size of the Brillouin zone. Physically, the particle can transfer its momentum to the underlying lattice as long as the transferred momentum is equal to one of the reciprocal lattice vectors, and therefore, the momentum of the particle is only defined within the bounds of the first Brillouin zone. This surprising result, which occurs only due to the presence of the lattice, implies that the velocity of the particle {\it oscillates} about zero in the presence of a {\it constant force}, and is called Bloch oscillations. For electronic materials in constant electric field, the time required for the requisite change of momentum is given by $t_B= \Delta p/(qE)\sim\hbar/(qE d)$ where $q$ is the electronic charge, $E$ is the applied, constant electric field, and $d\sim$ few \AA\, is the lattice constant. For typical fields $E\sim 10^3$ V/m, this time is orders of magnitude longer than the typical time $t_s$ between electron-lattice scatterings, $t_B\sim 10^{-8}$ sec $\gg t_s\sim 10^{-14}$ sec~\cite{bz,mermin}. Therefore, although long predicted in electronic systems, Bloch oscillations have not been and are unlikely to be observed in them. In addition, due to the large number of lattice sites, the effects of boundary on Bloch oscillations cannot be explored in electronic materials. Since there is no interaction of light with the dielectric and therefore no scattering that can randomize the transverse momentum of a wave packet in a lattice of waveguides, they provide an ideal platform to study Bloch oscillations and other energy-band related quantum phenomena in finite lattices where boundary effects can be prominent~\cite{Peschel1998}.  

% Bloch oscillations and edge effects.
\begin{figure*}[htbp]
\begin{center}
\includegraphics[width=\columnwidth]{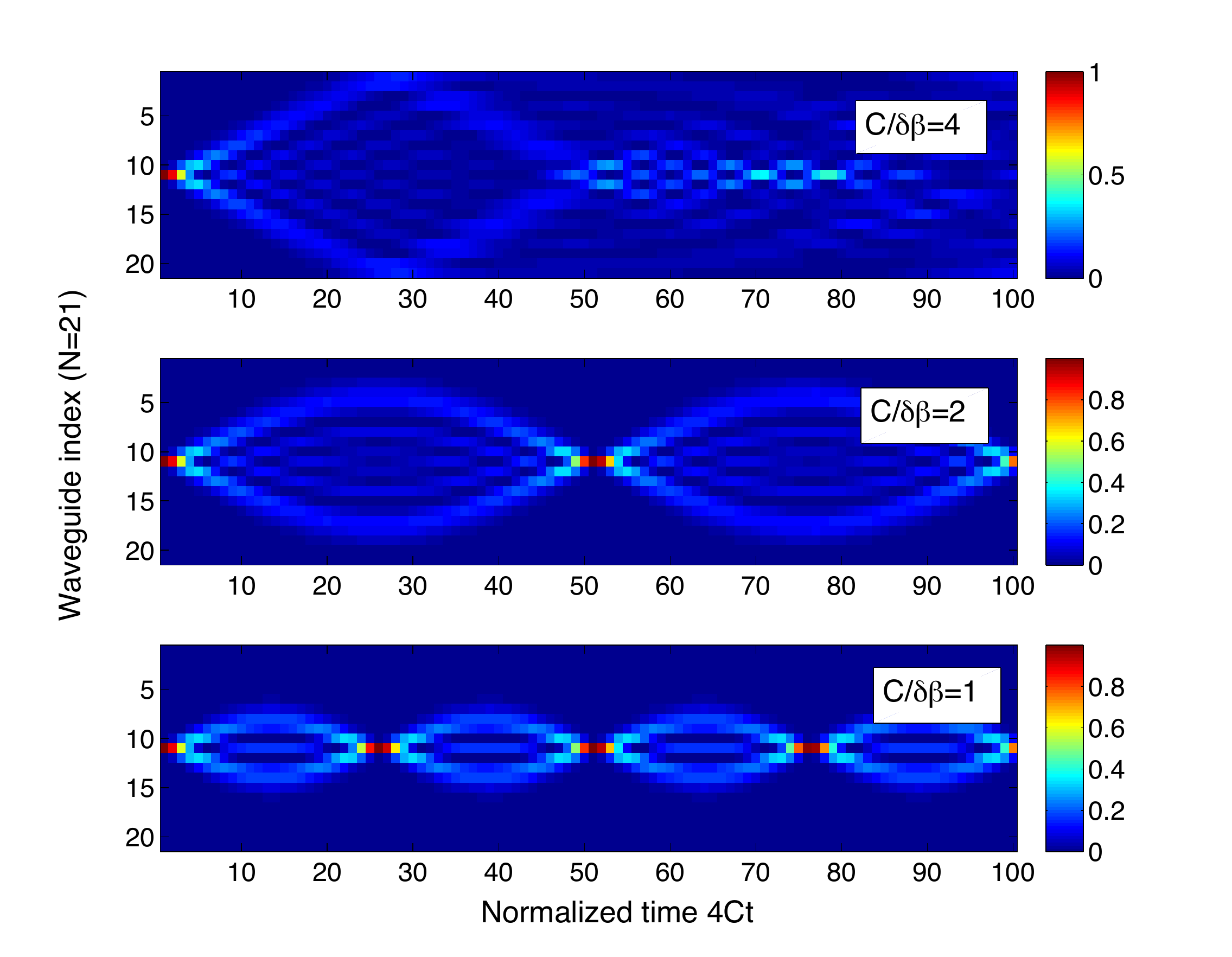}
\includegraphics[width=\columnwidth]{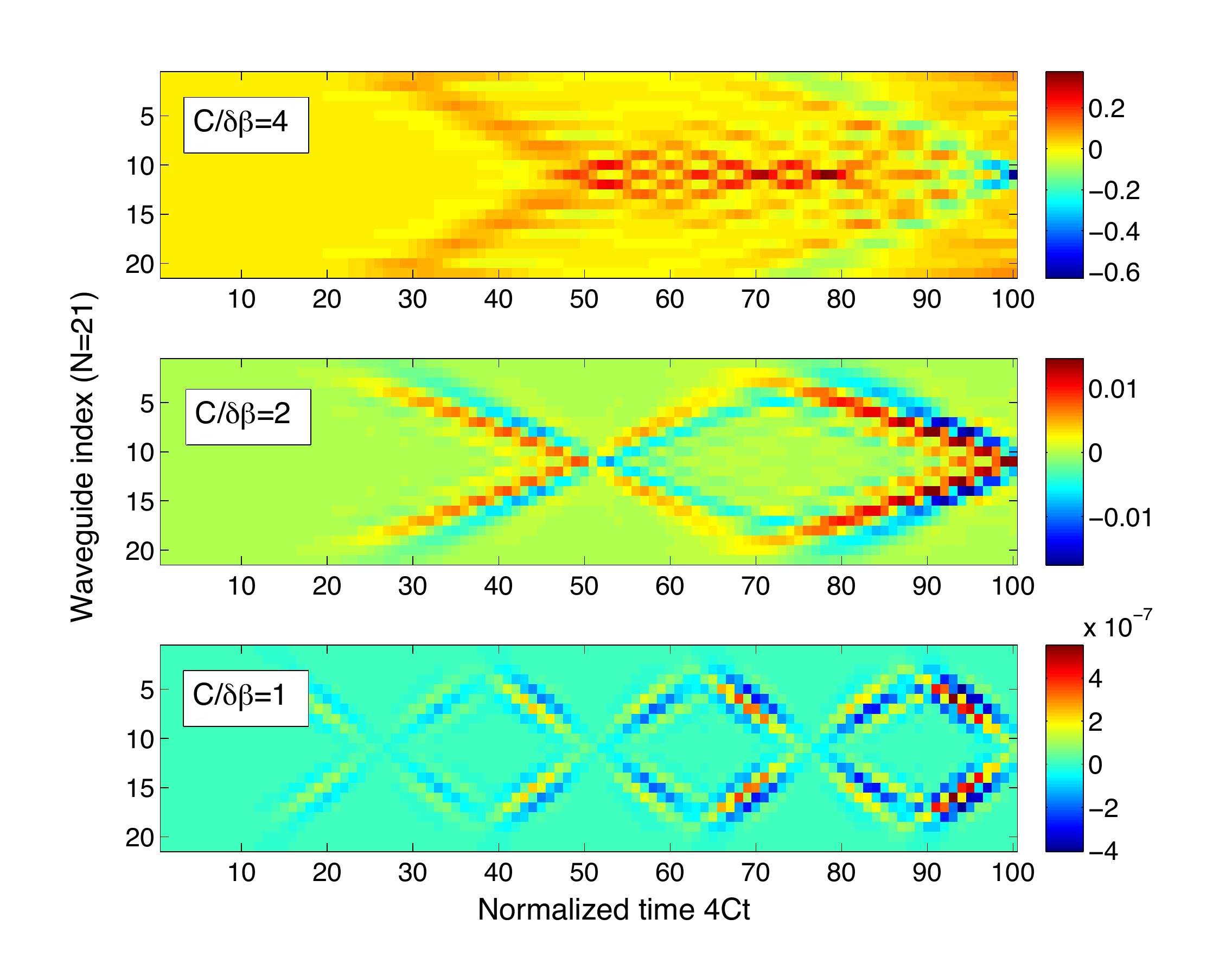}
\caption{The left-hand column shows the exact intensity $I(p,t)$ numerically obtained for a finite array of $N=21$ waveguides with a linear potential gradient $\delta\beta/C$, and the initial wave packet localized at the central site. The horizontal axis denotes time normalized in the units of $t_0=1/(4C)$. Bottom panel shows that for $\delta\beta/C=1$, the wave packet expands and contracts with period $T\propto 1/(\delta\beta)$. For a smaller gradient, $\delta\beta/C=0.5$ (center panel) the period of oscillation doubles and so does the maximum transverse extent of the wave packet. When $\delta\beta/C=0.25$ (top panel) the edge-reflection effects destroy the Bloch oscillations, although the intensity profile continues to remain symmetric about the center site, $I(p,t)=I(N+1-p,t)$. The right-hand column shows the corresponding differences $\Delta I(p,t)$ between the exact solution for a finite array and the analytical result for an infinite array. Note that, on average, $\Delta I(p,t)$ increases with time, but becomes appreciable only after the ballistically expanding wave packet has reached the boundaries.}
\label{fig:bloch}
\end{center}
\vspace{-5mm}
\end{figure*} 
To this end, we consider the waveguide array with a linear ramp in the on-site potential given by $\beta_j=\beta_0+\delta\beta j$ with $\delta\beta/\beta_0\ll 1$. Since $\beta_0$ only shifts the zero of the energy spectrum, we will ignore it in the subsequent treatment. This system is created by using variable-width waveguides with variable spacing between them to ensure constant tunneling and a linear gradient with $\delta\beta/\beta_0\sim 10^{-4}$~\cite{Peschel1998}. The equation of motion for the electric-field creation operator is given by $i\hbar\partial a^\dagger_j/\partial t=[H, a^\dagger_j]$ and reduces to
\begin{equation}
\label{eq:sch}
\frac{\partial a^\dagger_{j}}{\partial t}=+i(\beta_0+j\delta\beta) a^\dagger_{j}-iC(a^\dagger_{j+1}+a^\dagger_{j-1}),
\end{equation}
where one of the tunneling terms is absent when the site index $j$ corresponds to the first or the last waveguide in an array with $N$ waveguides. In the limit $N\rightarrow\infty$, this equation can be exactly solved by using Fourier transform~\cite{Thompson2010} and we get the following expression for the time-evolution operator $G(t)=\exp[-i Ht/\hbar]$ in the site-index space, 
\begin{eqnarray}
\label{eq:gf}
a^\dagger_j(t) & = &\sum_{m=-\infty}^{\infty} G_{jm}(t) a^\dagger_m(0),\\
\label{eq:gfsite}
 G_{jm}(t) & = & \exp\left[i(\beta_0+\delta\beta) t+\frac{i(j-m)(\delta\beta t-\pi)}{2}\right]\nonumber \\
& \times & J_{j-m}\left[\frac{4C}{\delta\beta}\sin\left(\frac{\delta\beta t}{2}\right)\right].
\end{eqnarray} 
Note that as the potential gradient vanishes, $\delta\beta\rightarrow 0$, we recover the propagator for a uniform lattice with bandwidth $\Delta_B=4\hbar C$. The time-evolution operator allows us to obtain the time and site-dependent intensity for an arbitrary normalized initial state $|\psi(0)\rangle=\sum_m \alpha_m a^\dagger_m(0)|0\rangle$,
\begin{equation}
\label{eq:int}
I(p,t)=|\langle p|\psi(t)\rangle|^2=|\sum_j \alpha_j G_{jp}(t)|^2,
\end{equation}
where the sum of weights is unity, $\sum_m |\alpha_m|^2=1$. If the initial input is confined to a single waveguide, $\alpha_m=\delta_{m,m_0}$, the intensity profile becomes 
\begin{equation}
\label{eq:blexact}
I_a(p,t)=J_{p-m_0}^2\left[\frac{4C}{\delta\beta}\sin\left(\frac{\delta\beta t}{2}\right)\right]. 
\end{equation}

This analytical result for the site and time-dependent intensity has the following features: It is symmetrical about the initial wave packet location; it is periodic in time with a period given by $T=2\pi/\delta\beta$; its maximum spread occurs at time $t=T/2$ and is determined by the ratio of the nearest-neighbor tunneling to the potential gradient $C/\delta\beta$. We emphasize that this result is valid only for an infinite array where the effects of boundaries can be ignored. On the other hand, since the tight-binding Hamiltonian Eq.(\ref{eq:tbh}) for a finite array corresponds to a finite, tri-diagonal, Hermitian matrix, one can obtain the time-evolved wave function $\langle p|\psi(t)\rangle$ exactly by straightforward numerical evaluation of the time-evolution operator $G(t)$. 

The left-hand panels in Fig.~\ref{fig:bloch} show the numerically obtained intensity $I(p,t)$ for an $N=21$ waveguide array with initial input in the central waveguide $m_0=(N+1)/2=12$; we use $t_0=\hbar/\Delta_B$ as the unit of time. For $\delta\beta=C$ (bottom panel), the period of Bloch oscillations is given by $T/t_0=8\pi C/\delta\beta=8\pi$ and the maximum spread of intensity is small compared to the size of the array. When $\delta\beta=0.5 C$ (center panel), the period is doubled, $T/t_0=16\pi$ and so is the vertical maximum spread of intensity. For $\delta\beta=0.25 C$ (top panel), the estimated wave packet spread is greater than the size of the array, and the open boundaries destroy Bloch oscillations although the intensity profile continues to remain symmetric about the center of the array. The right-hand panels in Fig.~\ref{fig:bloch} show the difference between numerically obtained intensity profile and the analytical result that is valid only for an infinite array, $\Delta I(p,t)=I(p,t)-I_a(p,t)$. When $\delta\beta=0.25C$ (top panel), the wave packet reaches the boundaries and thus the difference between the exact solution and the analytical result is the greatest, although we point out that this difference becomes appreciable only after the ballistically expanding wave packet has reached the array boundaries. When $\delta\beta=0.5C$ (center panel), the maximum intensity difference is approximately 1\% of the total intensity, although it increases with subsequent reflections from the boundaries of the finite array. When $\delta\beta=0.25C$ (bottom panel), the intensity difference $\Delta I$ is essentially zero. Thus, although the analytical result for the site and time dependent intensity is ideally applicable only for an infinite array, it accurately describes the dynamics of a finite array as long as the maximum spread of the wave packet does not detect the array boundaries.

% Ratchet motion with two-site initial state. 
\begin{figure*}[htbp]
\begin{center}
\includegraphics[width=\columnwidth]{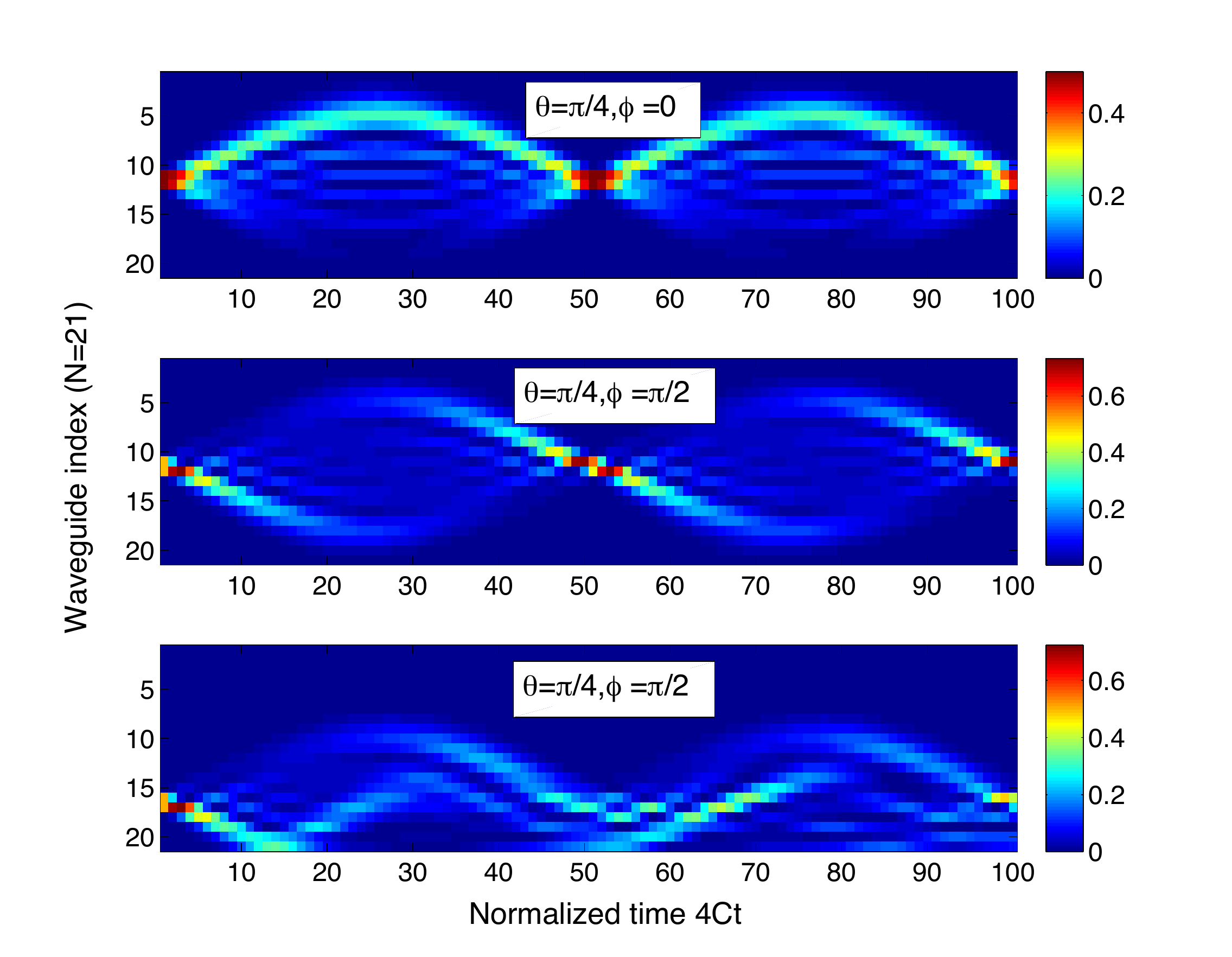}
\includegraphics[width=\columnwidth]{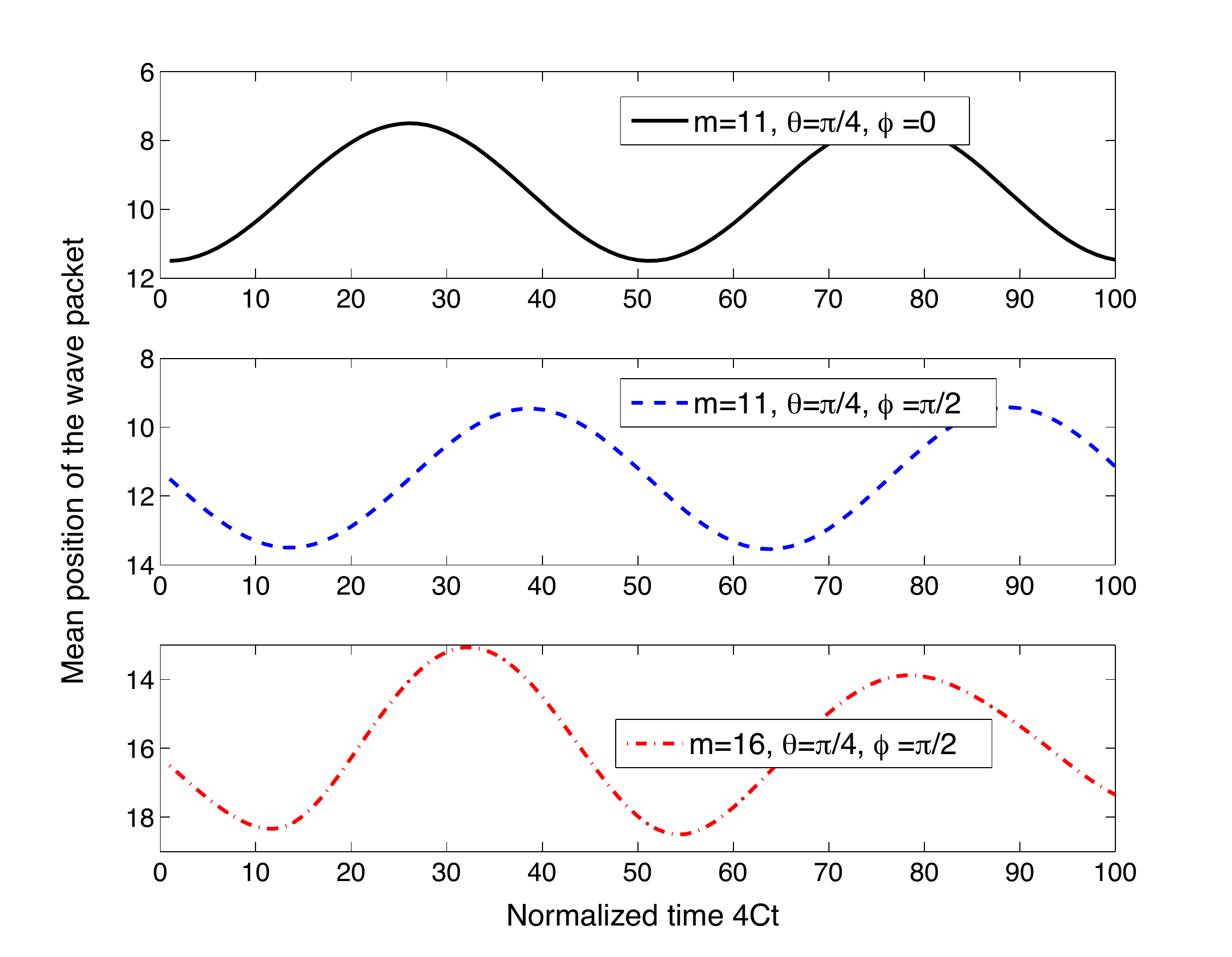}
\caption{Photonic transport in an $N=21$ array with linear gradient $\delta\beta/C=0.5$, initial state $|\psi(0)\rangle=\cos\theta|m_0\rangle+ \sin\theta e^{i\phi} |m_0+1\rangle$, and $\theta=\pi/4$. The left-hand column shows intensity profiles $I(p,t)$. When $m_0=11, \phi=0$ (top panel), the initially symmetric intensity profile shifts its weight towards the low potential region, whereas when $\phi=\pi/2$ (center panel), the weight oscillates from center to the high-potential region to the low-potential region. When $m_0=16, \phi=\pi$ (bottom panel) the wave packet weight starts to shift towards the high potential region, but the Bloch oscillations are destroyed due to reflections at the boundary. The right-hand column shows corresponding mean positions $j_\mathrm{mean}(t)$. For $m_0=11$ (top and center panels), the edge-effects are negligible and $j_\mathrm{mean}(t)$ oscillates with period $T=4\pi/\delta\beta$, consistent with Eq.(\ref{eq:mean}); when $m_0=16$ (bottom panel), the edge effects change this periodic behavior.} 
\label{fig:ratchet}
\end{center}
\vspace{-5mm}
\end{figure*}
The symmetrical intensity distribution in Bloch oscillations seen in Fig.~\ref{fig:bloch} is because all momenta within the Brillouin zone have equal weight in an input state that is localized to a single site. Next, we consider an initial state that is localized to two adjacent waveguides, with a phase difference $\phi$ between the two, $\alpha_m=\cos\theta\delta_{m0}+\sin\theta e^{i\phi}\delta_{m1}$. The analytical result for the site- and time-dependent intensity is given by 
\begin{eqnarray}
I(p,t) & = & \cos^2\theta J^2_p( \tau) + \sin^2\theta J^2_{p-1}(\tau) \nonumber\\
& - & \sin2\theta J_p(\tau)J_{p-1}(\tau)\sin(\phi-\delta\beta t/2),
\label{eq:ratchet}
\end{eqnarray}
where $\tau(t)=(4C/\delta\beta)\sin(\delta\beta t/2)$ and the last term in the intensity arises as a result of the interference between the two inputs. To quantify this interference, we consider the time-dependent average and standard deviation of the position, which, for an infinite array, can be simplified to 
\begin{eqnarray}
\label{eq:mean}
j_{\mathrm{mean}}(t) & = &\sum_m m I(m,t)  =  \sin^2\theta \nonumber \\ 
&+  & \sin 2\theta\frac{\tau}{2}\sin(\phi-\delta\beta t/2).\\
\label{eq:dev}
j^2_{\mathrm{std}}(t) & = & \sum_m m^2 I(m,t) =\sin^2\theta + \tau^2/2 \nonumber\\
& + & \sin 2\theta \frac{\tau}{2}\sin(\phi-\delta\beta/2).
\end{eqnarray}
Note that when the input is only confined to the central, zeroth waveguide, $\sin\theta=0$, we recover $j_{\mathrm{mean}}(t)=0$ and $j^2_{\mathrm{std}}=\tau^2/2$, and when the light is completely confined to the first waveguide, $\sin\theta=1$, we obtain the expected results. At small times, since the function $\tau(t)\approx 2Ct$, Eqs.(\ref{eq:mean})-(\ref{eq:dev}) imply that the mean position and its standard deviation both change linearly with time except when $\phi=\{0,\pi\}$; in those two cases, they change quadratically with time.  At large times, the mean position and standard deviation both oscillate due to the periodic nature of the function $\tau(t)=\tau(t+4\pi/\delta\beta)$. These results are only valid for an infinite array and, as we have seen earlier, they remain applicable to a finite array only if the maximum spread of the wave packet is smaller than the size of the array. 

Figure~\ref{fig:ratchet} shows the effects of the relative phase $\phi$ and initial wave packet on the intensity profile $I(p,t)$ (left-hand panels) and the mean position $j_{\mathrm{mean}}(t)$ (right-hand panels) for an $N=21$ waveguide array with $\delta\beta=0.5C$, period $T/t_0=16\pi$, and equally distributed weight on the adjacent sites, $\theta=\pi/4$. In the left-hand column, the top panels shows the asymmetrical intensity profile that results from an initially symmetric state $|\psi(0)\rangle=(|m_0\rangle+|m_0+1\rangle)/\sqrt{2}$ with $m_0=11$. The center panel shows corresponding intensity profile for $|\psi(0)\rangle=(|m_0\rangle+ i|m_0+1\rangle)/\sqrt{2}$, with $m_0=11$, where the asymmetry in the intensity profile switches direction with time. Both of these numerically obtained results are virtually identical with those obtained from Eq.(\ref{eq:ratchet}) that is valid for an infinite array. The bottom panel shows that the same wave function, with $m_0=16$, gives rise to an aperiodic intensity profile due to the presence of the boundary. The right-hand panels in Fig.~\ref{fig:ratchet} show corresponding mean position of the wave packet. When $\phi=0$ (top panel), the mean position is confined to the region of lower index of refraction and changes quadratically with time. When $\phi=\pi/2$ (center panel) we see that $j_{\mathrm{mean}}(t)$ oscillates about the initial mean position, and changes linearly with time at small times. The bottom panel shows that when the initial position is close to the boundary, the periodic behavior is destroyed due to the added interference with partial waves that are reflected from one edge of the array. Thus, the direction of the lateral photonic transport can be tuned by the relative phase difference $\phi$ between inputs at adjacent waveguides. 

%**************************************************************************************

\subsection{Continuum limit: non-relativistic particle}
\label{ss:cont}
In the last subsection, we considered the time evolution of a wave packet that is initially localized to one or two sites. Due to this extreme localization in real space, such a wave packet has components with all momenta (or equivalently, energies) across the entire bandwidth of the one-dimensional lattice. Due to the presence of these dimensionless momenta $-\pi<k\leq\pi$, the time evolution of the wave packet is dominated by quantum interference. On the other hand, by an appropriate choice of initial state that has energy components only near the bottom or the top of the cosine-band $E(k)=-2\hbar C\cos(k)$, one can mimic the behavior of a non-relativistic particle on a line segment. 

To formalize this mapping from a lattice to the continuum, let us consider lattice with sites $N\rightarrow\infty$ and site-to-site distance $d\rightarrow 0$ such that $Nd\rightarrow L$~\cite{dse}. We will choose a continuum co-ordinate system such that site $m=1$ maps to $x=-L/2$ whereas site $m=N$ maps to $x=+L/2$. In this limit, the nearest-neighbor tunneling term in Eq.(\ref{eq:sch}) translates into a spatial second-derivative with effective mass $m^*$ given by 
\begin{equation}
\label{eq:mass}
\frac{\hbar^2}{2m^*} = d^2\frac{\partial^2 E(k)}{\partial k^2}\bigg\vert_{k=0,\pi} = \pm d^2 \hbar C.
\end{equation}
Therefore, time evolution of an initial state $|\psi_e\rangle$ with components only near the bottom of the band, $k\sim 0$, in the presence of a linearly varying potential $V(x)=2\hbar\delta\beta x/L$ for $|x|\leq L/2$ should correspond to the time-evolution of a classical particle of mass $m^*=+\hbar/(2C d^2)$ in the presence of a constant force $F_0=2\delta\beta/L$ along the $-x$ direction. Borrowing the terminology from condensed matter physics, we call such a wave packet with positive effective mass electron-type or ``e-type''.  Equivalently, an initial state $|\psi_h\rangle$ with components near the top of the band, $k\sim\pm\pi$, corresponds to a classical particle with mass $m^*=-\hbar/( 2Cd^2)$ and will be called hole-type or ``h-type''.  We remind the reader that choosing purely real components  $\alpha_m$ for the initial wave packet ensures that the initial velocity of the classical particle is zero. 

% classical particle approximation motion.
\begin{figure}[htbp]
\begin{center}
\includegraphics[width=\columnwidth]{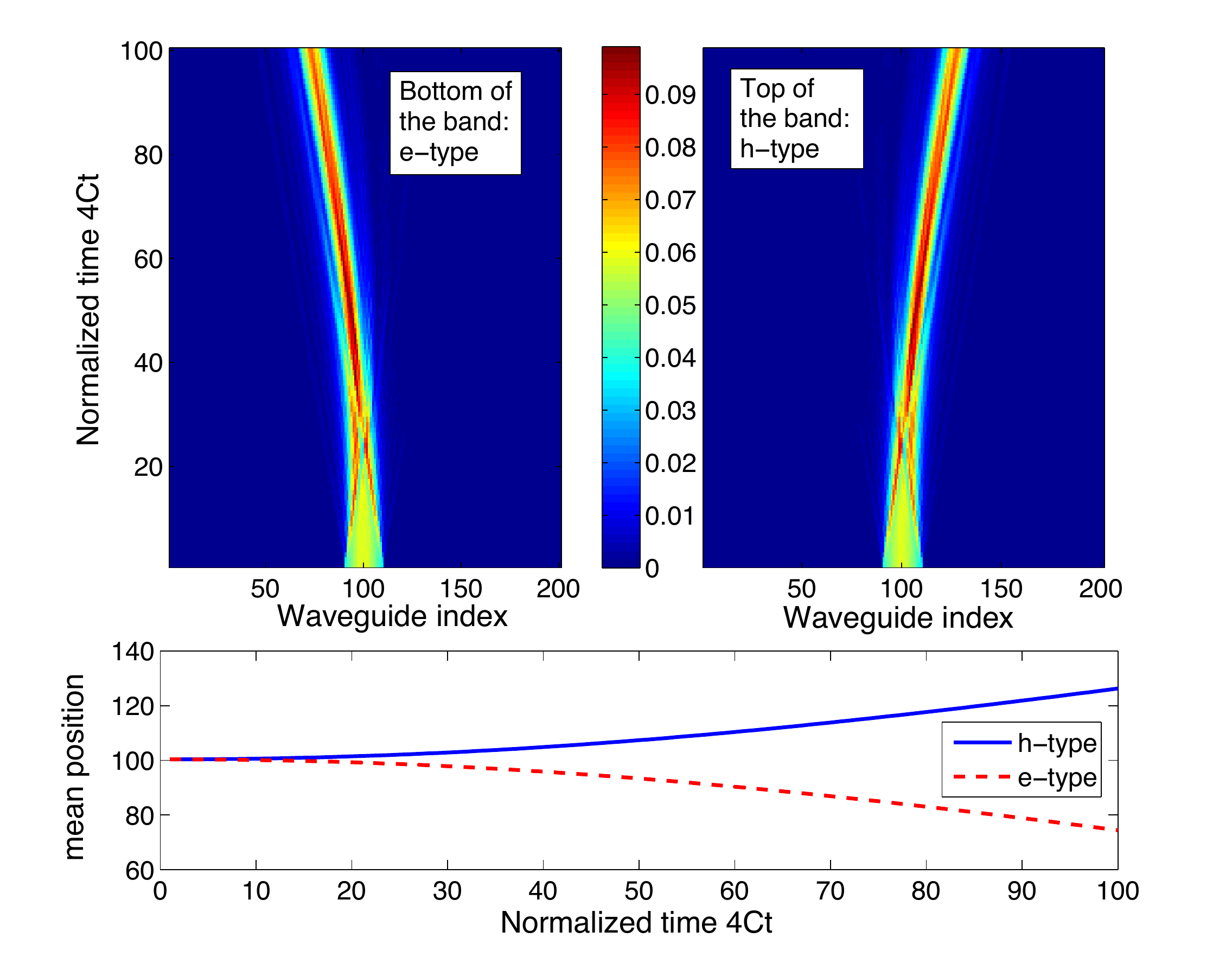}
\end{center}
\caption{Simulating a quasi-classical particle in a linear potential via an array with $N=201$ waveguides and potential gradient $\delta\beta/C=5$. The initial wave packet is spread across $M\sim N/10=20$ sites. The left-hand panel shows the intensity $I_e(p,t)$ for the ``e-type'' wave packet, which simulates a particle with positive effective mass, whereas the right-hand panel shows the corresponding result $I_h(p,t)$ for an ``h-type'' wave packet, which simulates a particle with negative effective mass. In contrast with the earlier results, here the wave packets (mostly) maintain their shape as they move towards lower or higher potential, respectively, in a parabolic manner. The bottom panel shows that the mean positions of the two  wave packets, $j_\mathrm{mean}(t)$, obtained from the time-dependent intensity distributions, follow the trajectory of a non-relativistic particle with constant acceleration and zero initial velocity.}
\label{fig:particle}
\end{figure}
Based upon this analysis, it follows that the average position of the wave packet $x(t)$ will satisfy 
\begin{equation}
\label{eq:quad}
x(t) = x(0) \mp \frac{F_0}{2|m^*|}t^2,
\end{equation}
where the negative sign is for an ``e-type'' wave packet, the positive sign is for an ``h-type'' wave packet, and $x(0)$ is the initial location of the wave packet. Figure~\ref{fig:particle} shows the numerically obtained results for time evolution of a wave packet in an array with $N=201$ waveguides and $\delta\beta=5C$. The top left panel shows the site and time-dependent intensity $I(p,t)$ of an ``e-type'' wave packet with initial Gaussian profile of size $M=N/10\gg1$ at the center of the array. We see that, in a sharp contrast with earlier results, the wave packet largely maintains its shape and moves toward the region with lower potential or, equivalently, smaller waveguide index, in a parabolic manner. The top right panel shows corresponding results for an identical ``h-type'' wave packet; it, too, maintains the shape, but moves towards larger waveguide index in a parabolic manner. We emphasize that in both cases, the external linear potential is identical; the opposite motions of the ``e-type'' and ``h-type'' wave packets arise due to their equal but opposite effective masses, and subsequent accelerations. These observations are quantified in the bottom panel where we plot the mean position of the wave packet, $j_{\mathrm{mean}}(t)$ as a function of normalized time for the ``e-type'' (dashed red) and ``h-type'' (solid blue) wave packets. It is clear that they follow Eq.(\ref{eq:quad}) where the magnitude of dimensionless acceleration is given by $|F_0/m^*d(4C)^2|=\delta\beta/(4CN)$, and matches the acceleration obtained from a quadratic fit to the data shown in the bottom panel. We emphasize that as the wave packet gets closer to the edge, the contribution from reflected partial waves increases and destroys its mapping onto a classical non-relativistic particle.  

These results show that a waveguide array with constant nearest-neighbor tunneling can be used to investigate properties of a quantum particle or a non-relativistic classical particle in an external potential. It also has the special property that the bandwidth of its corresponding Hamiltonian, $\Delta_B=4\hbar C$, does not depend upon the number $N\gg 1$ of waveguides in that array; this $N$-independence ensures the existence of the thermodynamic limit for such a lattice. However, as we discussed in the introduction, waveguide arrays offer the possibility of a site-dependent, nearest-neighbor tunneling $C_{k,k+1}=C_{k+1,k}=C(k)$. In the following subsection, we present the properties of arrays with such position-dependent tunneling profiles. 

%**************************************************************************************

\subsection{Arrays with site-dependent tunneling profiles}
\label{ss:parity}
For a finite array with $N$ waveguides and open boundary conditions, by judiciously choosing the distances $d_k$ between waveguides $k$ and $k+1$, any arbitrary tunneling profile $C(k)\geq 0$ can be created. For simple tunneling functions, the behavior of such an array can be easily deduced. For example, if $C(k)$ is a monotonically increasing function of site index $k$, then the average position of the quantum particle is shifted towards the end of the array with site index $N$. On the other hand, if $C(k)$ is a rapidly oscillating function of site index, $C(2k)\gg C(2k+1)$, then the $N$-site array is best understood in terms of $N/2$ weakly coupled dimers with tunneling profile $C(2k+1)$  where each dimer represents two adjacent waveguides with a strong tunneling $C(2k)$ between them~\cite{dimer}. In general, the tunneling profiles in both of these models break the parity-symmetry about the center of the array, $C(k)\neq C(N-k)$, and thus prefer one end of the array over the other. 

To maintain the equivalence between two ends of a finite, $N$-site array, we restrict ourselves to Hermitian tunneling profiles that obey $C(k)=C(N-k)$. In the simplest case, this constraint implies that the tunneling profile has either a single maximum or a single minimum at the center of the array. Therefore, we consider single-parameter tunneling functions 
\begin{equation}
\label{eq:talpha} 
C_\alpha(k)=C[k(N-k)]^{\alpha/2}=C_\alpha(N-k).
\end{equation}
When $\alpha>0$, the tunneling rate at the center of the array is $(N/4)^{\alpha/2}$ times larger than the tunneling near its edges, whereas when $\alpha<0$, the converse is true; when $\alpha=0$, we recover the constant-tunneling case. Since the tunneling amplitude $C(k)$ can be varied by a factor of hundred in a single material~\cite{naturereview,discreteo,alpha1,faithful}, establishing such tunneling profile constraints the size of the array to $(N/4)^{|\alpha|/2}\sim 100$ or, equivalently, $N\leq 10^4$ for $|\alpha|=1$, $N\leq 200$ for $|\alpha|=2$, and $N\sim 20$ for $|\alpha|=3$. These numbers show that it is feasible to fabricate waveguide arrays with a reasonable number of waveguides for tunneling profiles up to $|\alpha|\leq 3$. 

% properties of alpha-dependent spectrum. 
\begin{figure*}[htbp]
\begin{center}
\includegraphics[width=\columnwidth]{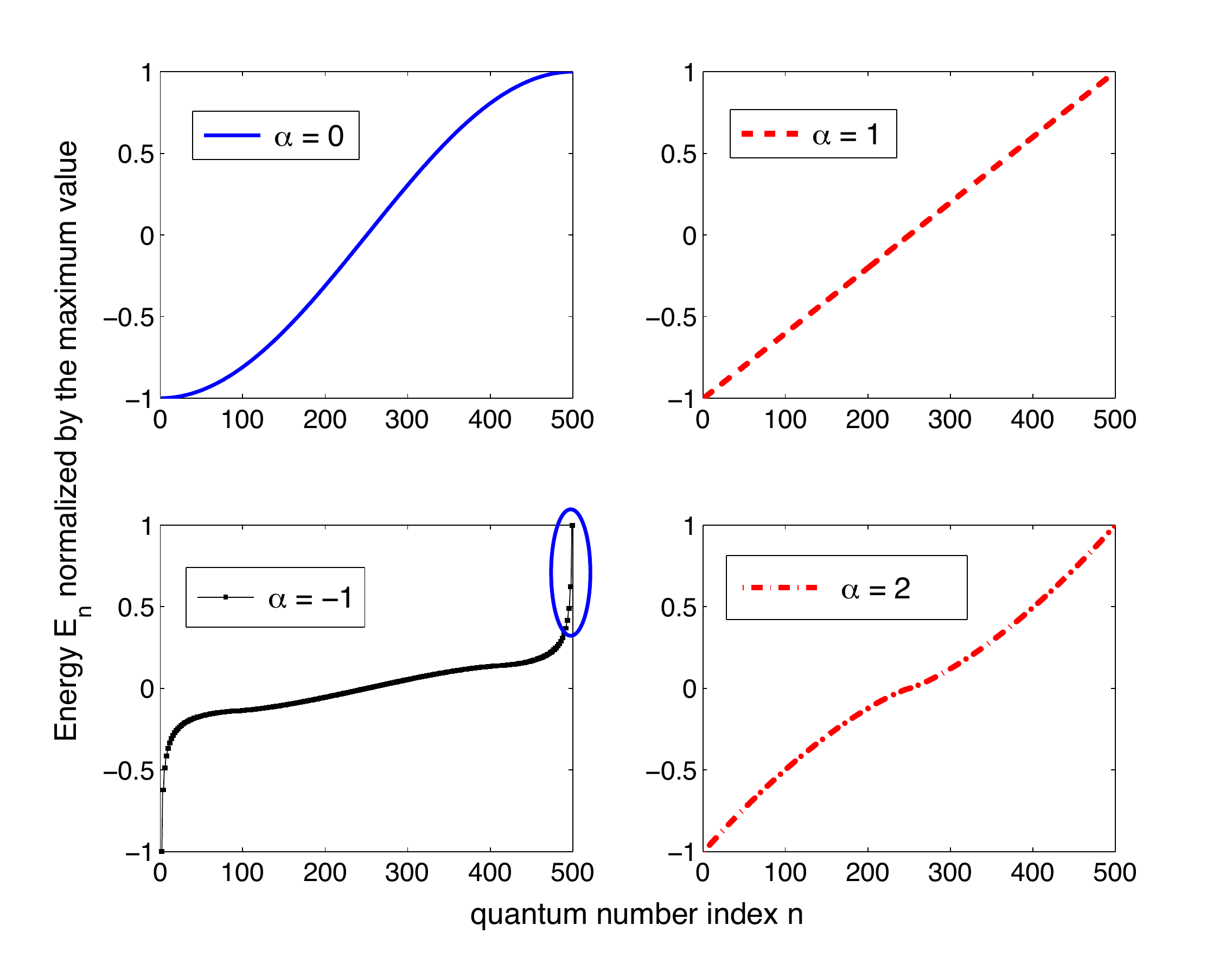}
\includegraphics[width=\columnwidth]{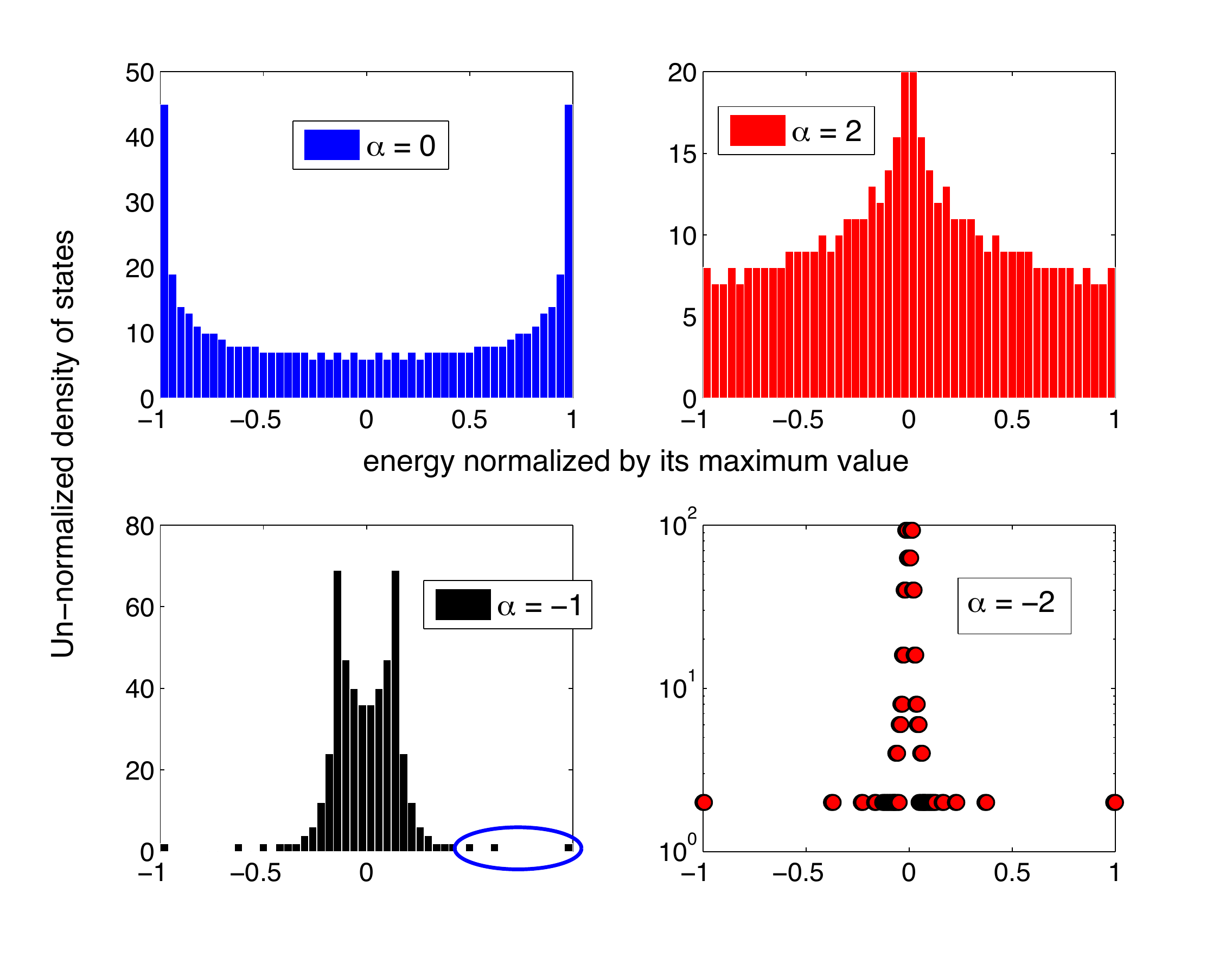}
\caption{Dimensionless energy spectra (left-hand four panels) and unnormalized density of states $D(\epsilon)$ as a function of dimensionless energy $\epsilon=E/E_\mathrm{max}$ (right-hand four panels) for Hamiltonian (\ref{eq:halpha}) with $N=500$ waveguides. The $\alpha=1$ spectrum is exactly linear, whereas for $\alpha=2$, it is linear near the edges. When $\alpha=-1$, the tunneling at the edge of the array is higher than that at its center, and the spectrum has discrete, localized states with energies near the band edges (the blue oval). On the right, when $\alpha=0$, $D(\epsilon)$ is maximum at $\epsilon=\pm 1$ whereas for $\alpha=2$, it is maximum at $\epsilon=0$. The quasilinear behavior of the $\alpha=2$ spectrum near the band edges is reflected in the flat $D(\epsilon)$ near $\epsilon=\pm 1$. For $\alpha<0$, the presence of discrete, localized states at the bottom and the top of the energy band is reflected in the finite, but vanishingly small, density of states away from the center of the band.}
\label{fig:spec}
\end{center}
\vspace{-5mm}
\end{figure*}
The Hamiltonian for such an $N$-site array is given by 
\begin{equation}
\label{eq:halpha} 
H_\alpha=\hbar\sum_{j=1}^{N-1}C_\alpha(j)\left[a_{j+1}^{\dagger}a_{j}+a_{j}^{\dagger}a_{j+1}\right].
\end{equation}
We remind the reader that when $\alpha\neq 0$, due to the loss of translational invariance, the eigenstates of the Hamiltonian are not labeled by momentum and, in general, it is not possible to obtain analytical solutions for the eigenvalues and eigenfunctions. The sole, notable exception is the case with $\alpha=1$, where analytical solutions for the eigenvalues and eigenfunctions are possible~\cite{gf,slalpha,ya}. One can, however, show that energy eigenvalues of $H_\alpha$ for any $\alpha$ occur in pairs $\pm E_n$ and that the corresponding eigenfunctions are related by a simple transformation~\cite{Joglekar2010a}. 

Figure~\ref{fig:spec} shows the typical properties of Hamiltonian $H_\alpha$, for an array with $N=500$ and $|\alpha|\leq 2$ obtained numerically. The left-hand four-panel figure shows the 
energy eigenvalues normalized by their respective maximum for $\alpha=\{0,1,2,-1\}$ (clockwise). For $\alpha=0$, we get the well-known cosine-band. When $\alpha=1$, we obtain a spectrum with equidistant energy eigenvalues, maximum eigenenergy $E_\mathrm{max}=(N-1)\hbar C$, and level spacing $\Delta E=2\hbar C$; for $\alpha=2$, the spectrum is linear near the band edges, with a flatter region in between. For $\alpha=-1$, the spectrum consists of a few localized states near the band edges (shown by the blue oval) along with a bulk of extended states~\cite{ya}. The four panels on the right-hand side show the unnormalized density of eigenstates $D(\epsilon)$, which provides a measure of number of eigenstates available in a small interval $\delta\epsilon$ around energy $\epsilon$ for $\alpha=\{0,2,-2,-1\}$ (clockwise). For $\alpha=0$, we recover the well-known result for a one-dimensional lattice with van-Hove singularities, signaled by a diverging $D(\epsilon)$, at the band edges~\cite{bz,vhs}. For $\alpha=1$, due to the equidistant energy levels, the density of states is a constant. When $\alpha=2$, the density of states has a maximum near zero energy, consistent with the small slope of the corresponding energy spectrum near $\epsilon=0$. When $\alpha=-1$, the $D(\epsilon)$ has two distinct features. The first is a two-peaked structure that represents the density of bulk, extended states; the second is the presence of discrete, localized states near the band edges (shown by the blue oval). When $\alpha=-2$, these features are preserved, but there are a number of localized states at different energies; note that the logarithmic vertical scale in this panel shows the distributed weight of such states. These results show that arrays with $\alpha$-dependent tunneling have widely tunable spectra. 

We define the energy bandwidth as $\Delta_\alpha(N)=E_\mathrm{max}-E_\mathrm{min}=2E_\mathrm{max}$. When $\alpha = 0$, the bandwidth is independent of the array size  for $N\gg 1$, $\Delta_{\alpha=0}(N)\rightarrow \Delta_B=4\hbar C$, whereas for $\alpha\neq 0$, the bandwidth depends upon the size of the array and is essentially determined by the maximum tunneling element in the array. Thus, $\Delta_\alpha(N)\sim N^\alpha$ for $\alpha >0$ and $\sim N^{-|\alpha |/2}$ for $\alpha <0$. In the following, we use inverse-bandwidth as the characteristic unit of time for an array with a given tunneling profile $\alpha$ and number of waveguides $N$,  $\tau_\alpha(N)=\hbar/E_\mathrm{max}=2\hbar/\Delta_\alpha(N)$. Thus, as $\alpha>0$ increases, the characteristic time $\tau_\alpha$ and the characteristic length $l_\alpha= c\tau_\alpha/n$ both decrease, where $c/n$ is the (constant) speed of light along the waveguide with index of refraction $n$. Thus, in a sample with a given physical length, long-time dynamics are easily observed as $\alpha$ increases, whereas short-time dynamics become accessible for $\alpha<0$~\cite{clinttunable}. 

Now we consider the time evolution of a wave packet in such an array. For an arbitrary initial state $|\psi(0)\rangle$, the time-evolved state is obtained by $|\psi(t)\rangle= G_\alpha(t) |\psi(0)\rangle$ where the time-evolution operator $G_\alpha(t)=\exp\left[-i H_\alpha t/\hbar\right]$ is obtained numerically. Since we have discussed the time-dependent intensity profiles of wave packets that are localized to a single or two sites in Sec.~\ref{ss:bloch}, here we choose a broad initial state that is equally distributed across all waveguides, $|\psi(0)\rangle=1/\sqrt{N}$. 

% clean alpha-dependent wave packet evolution.
\begin{figure}[htbp]
\begin{center}
\includegraphics[width=\columnwidth]{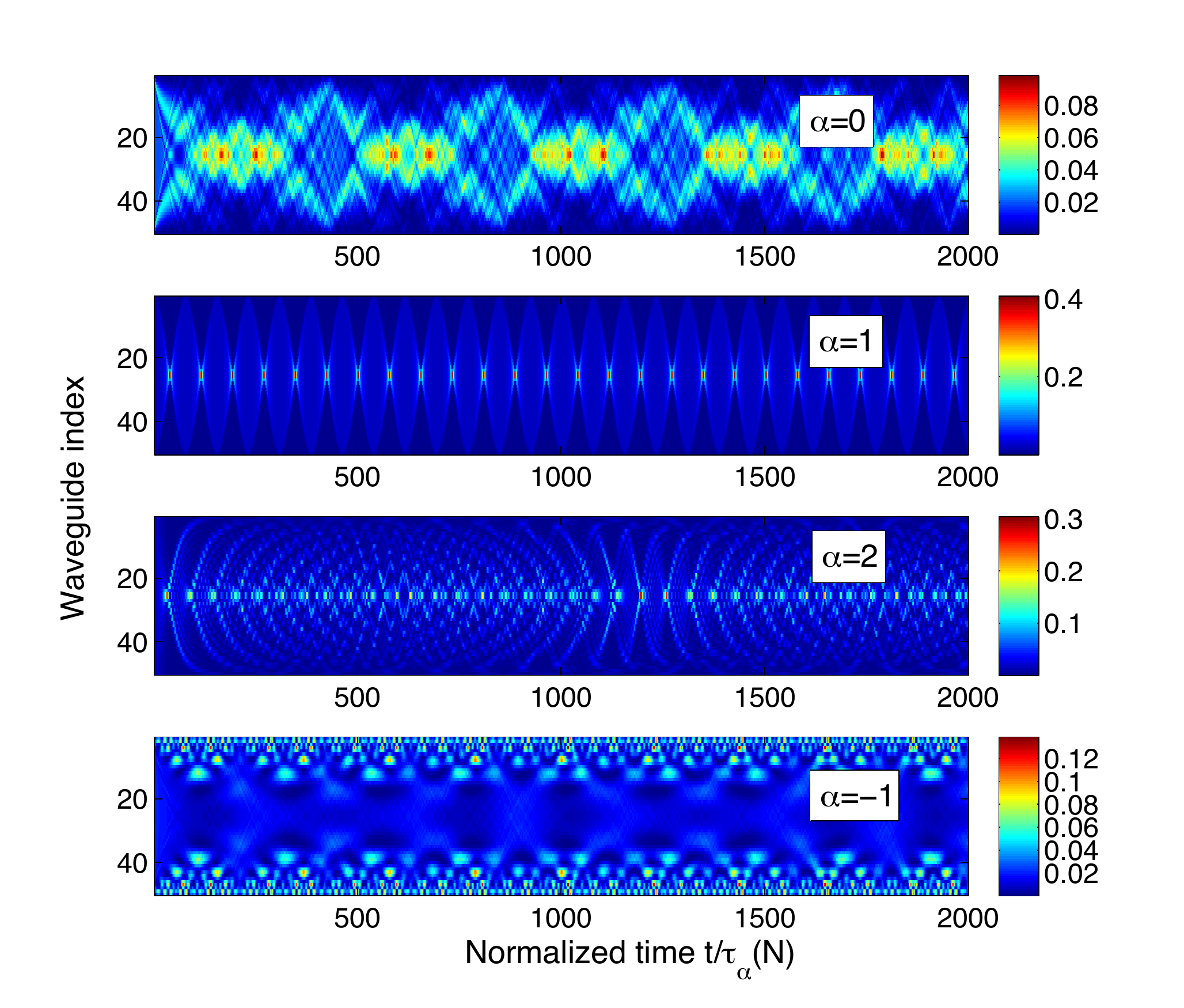}
\caption{$I(p,t)$ for a uniformly distributed initial state, $|\psi(0)\rangle=1/\sqrt{N}$; the horizontal axis denotes normalized time. For $\alpha=0$, we get larger intensity in the central region due to edge-reflection and interference. $\alpha=1$ shows periodic behavior due to the equally-spaced eigenvalues of the underlying Hamiltonian. When $\alpha=2$, the quasilinear energy spectrum and the edge-reflections contribute to the quasi-periodic larger intensity in the central region.  The bottom panel shows that for $\alpha=-1$, eigenstates localized at the two edges lead to a larger intensity at the edge instead of in the central region.}
\label{fig:alphaclean}
\end{center}
\vspace{-5mm}
\end{figure}
Fig.~\ref{fig:alphaclean} shows the intensity $I(p,t)=|\langle p| G_\alpha(t)|\psi(0)\rangle|^2$ in an $N=50$ array with $\alpha=\{0,1,2,-1\}$; the horizontal axis denotes time normalized by the $\alpha$- and $N$-dependent time-scale $\tau_\alpha(N)$. Note that, due to the symmetries of the Hamiltonian and the initial state, the intensity satisfies $I(p,t)=I(N+1-p,t)$ and that the average intensity per site is $I_a=0.02=1/N$. When the tunneling is constant, the effects of interference and reflection at the boundaries lead to a suppression of the intensity at the edges, and a modest enhancement, by a factor of five, near the center of the array ($\alpha=0$ top panel).  When $\alpha=1$ (second panel) the constant spacing between the energy levels implies that the intensity profile is periodic in time, $I(p,t)=I(p,t+\pi/C)$. In contrast to the constant tunneling case, we also observe that the maximum intensity at the center of the array is enhanced by a factor of 20. For $\alpha=2$ (third panel) due to the quasilinear nature of the energy spectrum, we see approximate reconstruction of the intensity profile, and the maximum intensity at the center is again significantly enhanced from its initial value. In all the  three cases, since the tunneling at the center is maximum, we see that the intensity profile $I(p,t)$, in general, is largest at the center of the array and reduces symmetrically on the two sides. The bottom panel in Fig.~\ref{fig:alphaclean} shows the intensity evolution for an array with $\alpha=-1$, which has localized eigenstates at the two ends of the array. In a sharp contrast with the earlier results, we see that $I(p,t)$ now shows symmetrical maxima near the two edges of the array, with a broad minimum near the central region. These results show that identical initial states give rise to strikingly different intensity profiles in tunable waveguide arrays with a position-dependent tunneling profiles. 

%**************************************************************************************

\subsection{Disorder induced localization}
\label{ss:disorder}
In the past three subsections, we have focused on the properties of waveguide arrays with constant or position dependent tunneling profiles and constant or linearly varying on-site potentials; we implicitly assumed that it was possible to fabricate a waveguide array with the exactly specified Hamiltonian. This is, of course, an approximation. In real samples, disorder is always present through variations in the tunneling amplitudes $\hbar C_{j,j+1}$ or on-site potentials $\hbar\beta_j$ in the tight-binding Hamiltonian, Eq.(\ref{eq:tbh}). The effect of such disorder on the transport properties of lattices was first investigated in the context of electronic systems~\cite{anderson1,leerama}, and then extended to classical waves~\cite{all1,all2,all3}. In one dimension, all eigenstates of a disordered Hamiltonian are exponentially localized in the limit of an infinite system size, $N\rightarrow\infty$ irrespective of the strength of the disorder $v_d$. This non-analytical result - exponential localization at infinitesimal disorder - is due to the subtleties associated with the order of limits $N\rightarrow\infty$ and $v_d\rightarrow 0$~\cite{1Dal1,1Dal2,1Dal3}. 

In a finite array of $N\sim 10^2$ coupled waveguides, {\it localization} refers not to an exponential localization of all eigenstates {\it \`{a} la} electronic systems, but rather to the development of a ``steady-state'' intensity profile $I(p)$ that contrasts the ballistic expansion and edge-reflection present in a clean system. The time required for the emergence of the steady-state profile is inversely proportional to the strength of the disorder. In another sharp contrast, the typical strength of disorder in (weakly conducting) electronic materials is $v_d\ll E_F$ whereas in waveguide arrays, the disorder strength can be comparable to the tunneling, $v_d\sim\hbar C$~\cite{Lahini2008,Thompson2010}. 

% dirty alpha-dependent wave packet evolution.
\begin{figure*}[htbp]
\begin{center}
\includegraphics[width=\columnwidth]{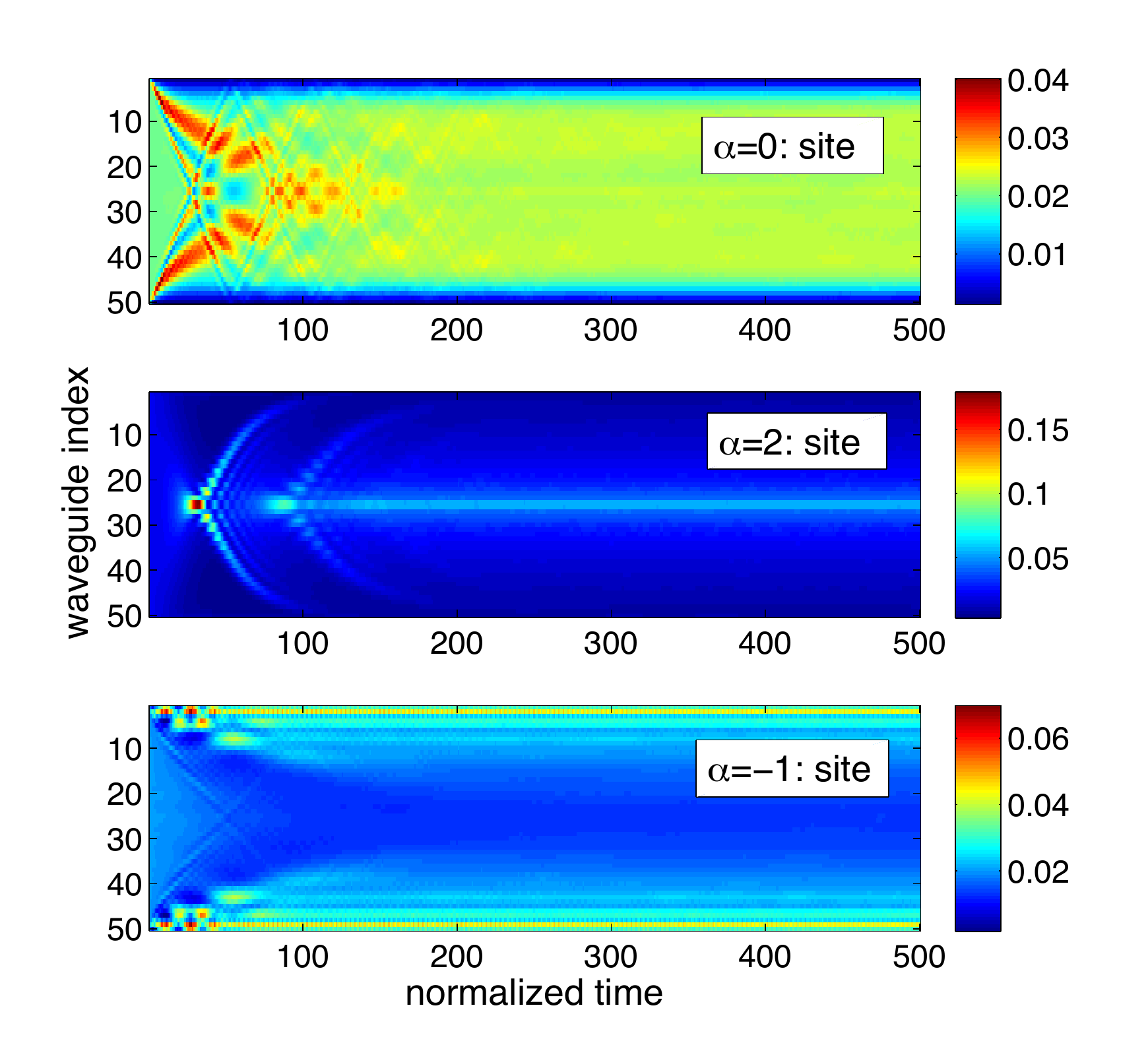}
\includegraphics[width=\columnwidth]{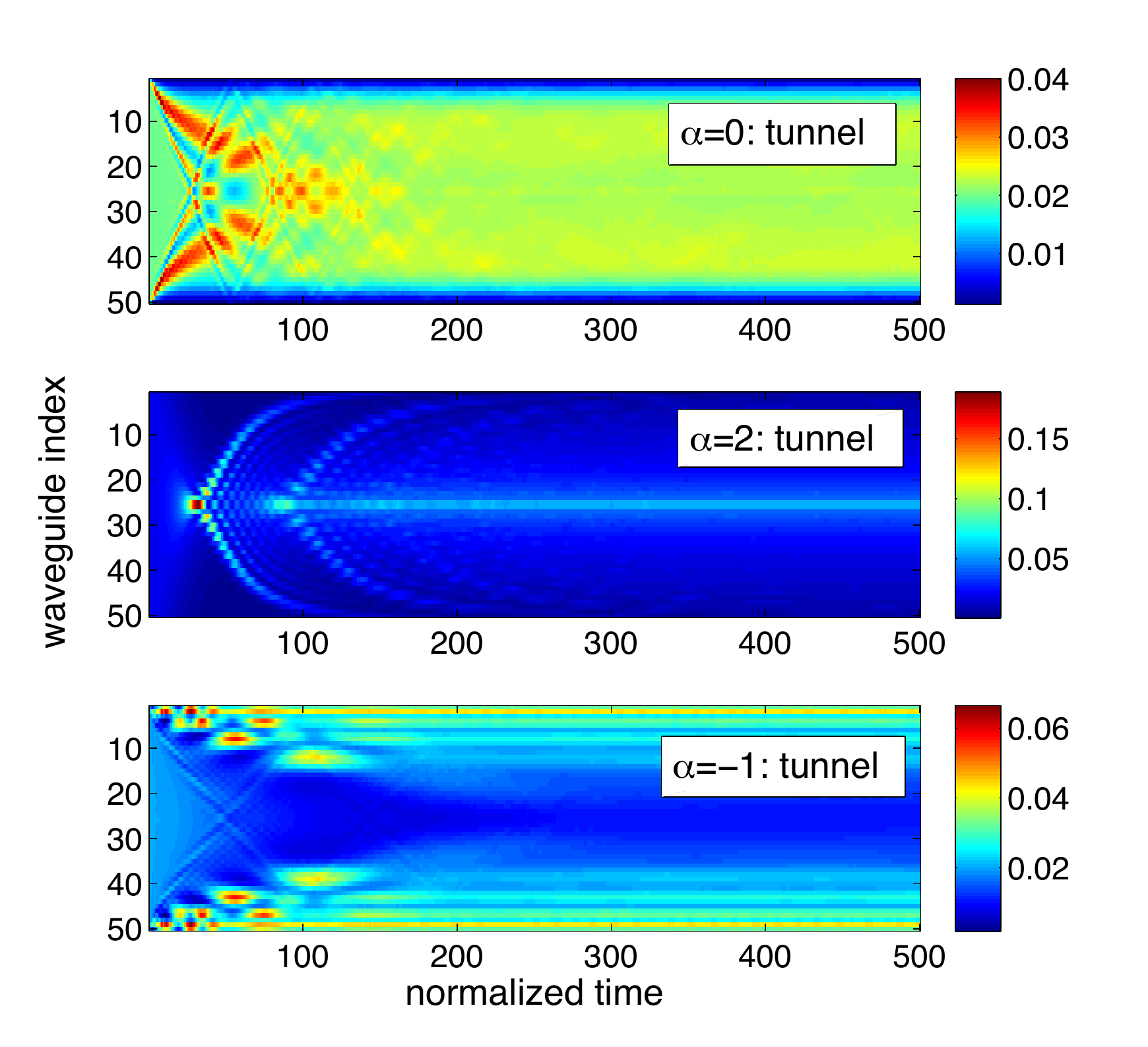}
\caption{Disorder averaged intensity profiles $\langle I(p,t)\rangle$ for a uniform initial state with site-disorder (left-hand column) and tunneling-disorder (right-hand column) in an $N=50$ array; the horizontal axes denote time normalized by the relevant time-scale $\tau_\alpha(N)$ and the two disorder strengths are equal, $v_{ds}=v_{dt}=0.1\Delta_\alpha$. In all cases, the interference pattern at small times is replaced by quasi steady-state intensity at large times. For $\alpha=0$ (top line) and $\alpha=2$ (center line), $\langle I(p)\rangle$ has a maximum near the central region, whereas for $\alpha=-1$ (bottom line) the intensity has multiple maxima near the two edges of the array. This emergence of steady state profiles shows that ``extended'' initial states also undergo disorder-induced ``localization'' as it is defined here.}
\label{fig:alphadisorder}
\end{center}
\vspace{-5mm}
\end{figure*}
In this subsection, we present the effects of disorder on the time-evolution of a uniform initial state. We consider two distinct disorders. The diagonal disorder randomly modulates the on-site potential $\hbar\beta_i\rightarrow \hbar\beta_i+v_i$ where $v_i$ is a random variable with zero mean and variance $v_{ds}$. The off-diagonal disorder randomly modulates the tunneling $\hbar C_i\rightarrow\hbar C_i+v_i$ where $v_i$ is a zero-mean random variable with variance $v_{dt}$. We use uniformly distributed random variables to ensure that the modulated tunneling rates remain strictly positive, although the results are independent of the type of distribution used as long as any such distribution has zero mean and identical variance~\cite{Thompson2010,karr2011}. The resultant intensity distribution is averaged over multiple $M\sim 10^4$ realizations to ensure that the final results are independent of the number of disorder realizations and the probability distribution of the site or tunneling disorder. Figure~\ref{fig:alphadisorder} shows the intensity profile $\langle I(p,t)\rangle$ for an array with $N=50$, uniform initial state, and $\alpha=\{0,2,-1\}$ where $\langle\cdots\rangle$ denotes disorder average. We remind the reader that the average intensity per site is $I_a=1/50$. The left-hand column has results for on-site disorder $v_{ds}$ and the right-hand column has results for the tunneling disorder of equal strength, $v_{ds}=v_{dt}=0.1\Delta_\alpha(N)$. The top line, $\alpha=0$, shows that for both disorders the initial interference pattern is replaced at later times by a steady state intensity that is suppressed at the edges. The center line, $\alpha=2$, shows the same qualitative behavior, but also shows slight difference between the the two intensity profiles, particularly at small times. The bottom line, $\alpha=-1$, shows steady-state profiles that have maxima near the two edges. In all cases, the differences between the left-hand and right-hand panels for a given tunneling profile $C_\alpha(j)$ decrease with increasing time, measured in units of $\tau_\alpha(N)$. 

% Steady-state alpha-dependent wave packet intensity.
\begin{figure}[htbp]
\begin{center}
\includegraphics[width=\columnwidth]{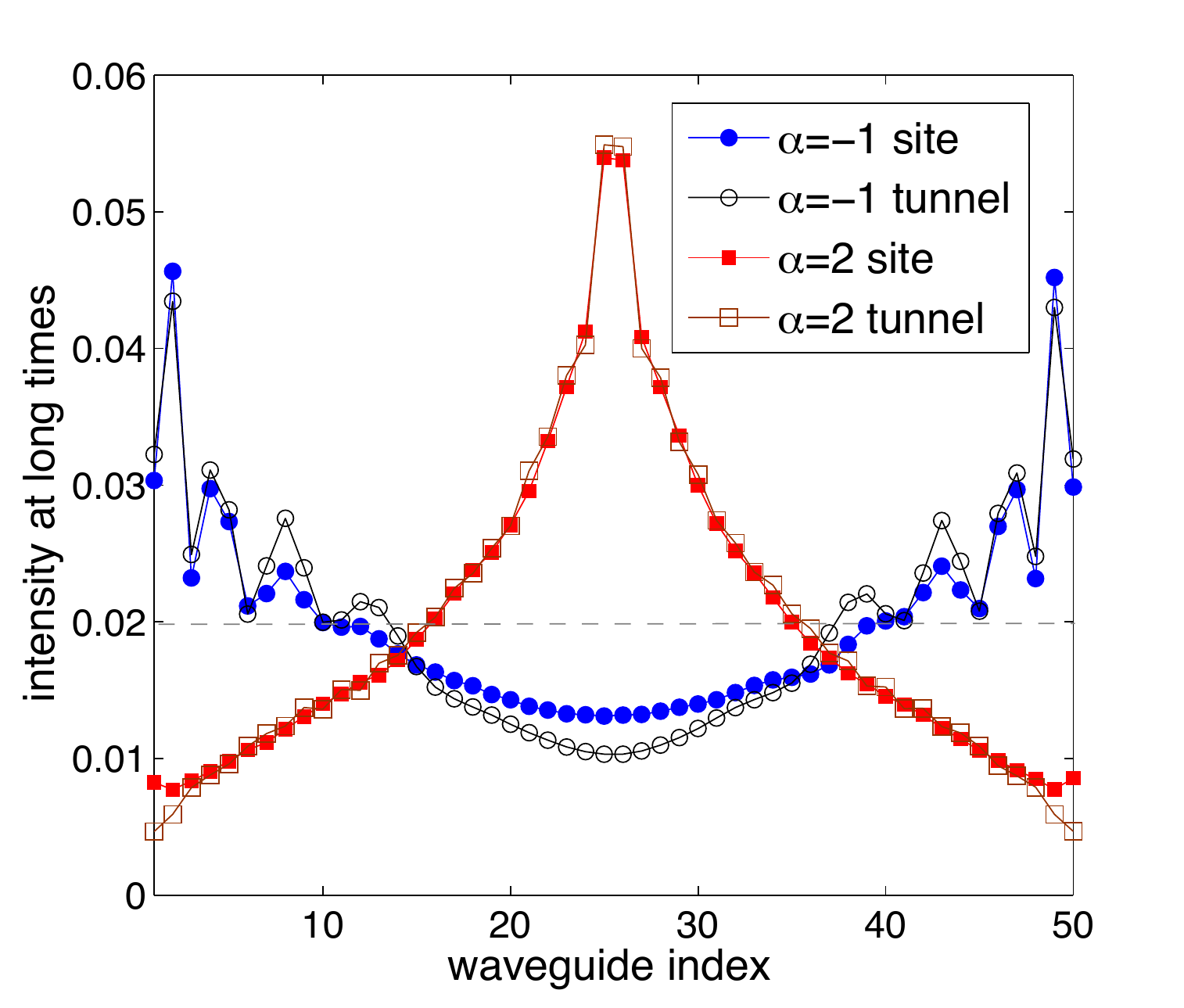}
\caption{Intensity profiles at $t/\tau_\alpha=500$ for an array with $N=50$, $\alpha=2$ (squares) and $\alpha=-1$ (circles), and on-site (solid symbols) or tunneling (open symbols) disorders of equal strength. Both disorders give identical disorder-averaged intensity profiles at sufficiently long times, and they are parity-symmetric about the center of the array.}
\label{fig:steady}
\end{center}
\end{figure}
Lastly, we compare the cross-section of the intensity profiles at $t/\tau_\alpha=500$ for the same array  with on-site disorder (solid symbols) and tunneling disorder (open symbols) of equal strength, $v_{ds}=v_{dt}=0.1\Delta_\alpha$. When $\alpha=-1$ (circles), the intensity profile shows a minimum at the center and multiple, symmetric maxima at the two edges, whereas for $\alpha=2$ (squares), the intensity is maximum at the center and monotonically decays away from it. Note that the multiple maxima near the two edges show up as striations in the intensity profiles for $\alpha=-1$ in Fig.~\ref{fig:alphadisorder}. The (gray) dashed line shows the average intensity $I_a=1/N=0.02$ per site. We point out that the intensity profiles for on-site and tunneling disorders coincide with each other at sufficiently long times, although the time required for such a match depends upon the tunneling profile $\alpha$ and the initial state. For example, Fig.~\ref{fig:steady} shows virtually identical intensity profiles for $\alpha=2$, whereas for $\alpha=-1$, the intensity suppression due to the tunneling disorder (black open circles) is larger than that by the on-site disorder (blue solid circles). It is also worth emphasizing that the disorder-averaged intensity profile recovers the underlying parity-symmetry shared by the clean Hamiltonian and the initial state, $\langle I(p,t)\rangle=\langle I(N+1-p,t)\rangle$.  

% Mirror-symmetric localization intensity.
\begin{figure}[bhtp]
\begin{center}
\includegraphics[width=\columnwidth]{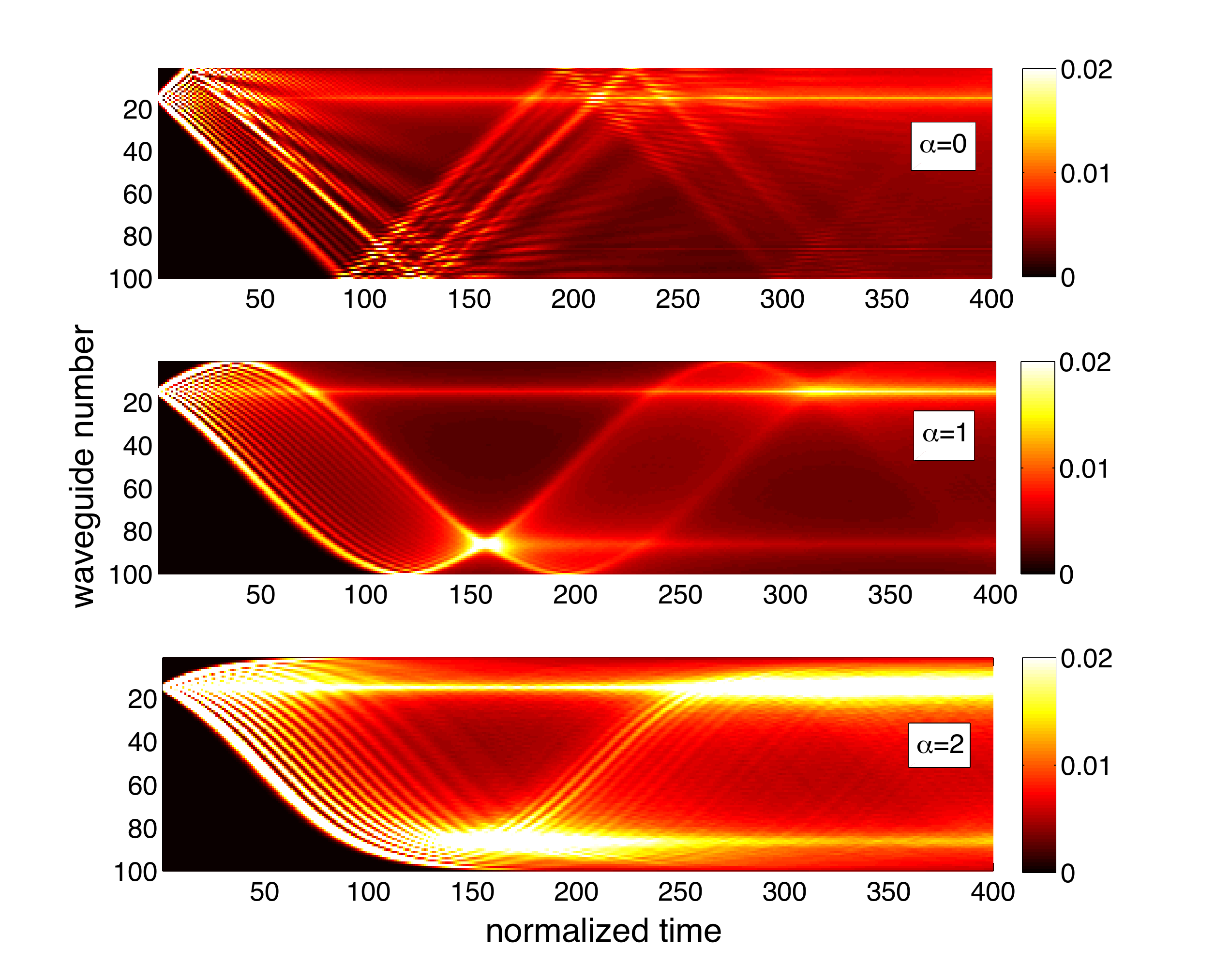}
\caption{Intensity $\langle I(p,t)\rangle$ in an array with $N=100$, a weak disorder $v_{ds}/\Delta_\alpha=0.05$, and initial wave packet $|m_0\rangle$ with $m_0=15$. Top panel shows that for constant tunneling, the steady-state intensity profile is maximum at $m_0$, with exponential decay on the two sides. The center ($\alpha=1$) and bottom ($\alpha=2$) panels show that, at short times, the wave packet partially reconstructs at the mirror-symmetric site $\bar{m}_0=N+1-m_0=86$.  Thus, in sharp contrast with the traditional localization, $\alpha=0$ case, the steady-state intensity profiles for $\alpha\geq 1$ have a two peaks, one at the initial wave packet location and the other at its parity-symmetric counterpart. Note that in all cases, the average intensity is $I_a=0.01=1/N$ and thus, the localization enhancement is only by a factor of two.}
\label{fig:mirrorloc}
\end{center}
\vspace{-5mm}
\end{figure}
We end this section with another phenomenon due to the parity-symmetric tunneling profile in a finite array of waveguides. Fig.~\ref{fig:mirrorloc} shows the time-and site-dependent intensity evolution in an array with $N=100$ waveguides, a small on-site disorder $v_{ds}/\Delta_\alpha=0.05$, and tunneling profiles with $\alpha\geq 0$. The initial wave packet is {\it localized at a single site} $m_0=15$. The top panel ($\alpha=0$) shows that $I(p,t)$ changes from interference-dominated behavior at short times to disorder-dominated steady-state behavior at longer times; the steady-state intensity is maximum at site $m_0$ and decays exponentially with distance from $m_0$~\cite{Lahini2008}. The center panel ($\alpha=1$) shows that, at short times, the wave packet partially reconstructs at the parity-symmetric site $\bar{m_0}=N+1-m_0$. The steady-state intensity profile in this case has two peaks, at $m_0$ and $\bar{m}_0$, and their relative weights are tuned by the disorder strength and the distance between the two peaks. The bottom panel ($\alpha=2$) shows a qualitatively similar result. Thus, a position-dependent, parity-symmetric tunneling in a finite array of waveguides leads to effective localization at two waveguide locations, even if the initial wave packet is introduced in a single waveguide~\cite{clintdisorder}.

%**************************************************************************************
%**************************************************************************************q

\section{Non-Hermitian, $\mathcal{PT}$-symmetric Models}
\label{sec:pt}
In the last section, we only considered Hermitian Hamiltonians, Eq.(\ref{eq:tbh}), which modeled 
waveguides that have no loss or amplification of the input signal. The ubiquitous losses that are 
present in real waveguides are phenomenologically taken into account by adding a negative imaginary part to the real eigenvalues of the Hermitian Hamiltonian, $E_n\rightarrow E_n-i\Gamma_n$~\cite{griffith,leggett}. This imaginary part $\Gamma_n>0$ leads to an exponential decay of the total intensity and therefore represents dissipation, absorption, or friction~\cite{wen}. Nominally, if we assign a {\it positive imaginary part} to the energies of a Hermitian Hamiltonian, $E_n\rightarrow E_n+i\Gamma_n$ with $\Gamma_n>0$, the total intensity of an initially normalized wave packet will increase, and will therefore represent gain or amplification. Such a phenomenological model breaks down at long times, when the power required to maintain the exponential intensity increase cannot be supplied by the ``reservoir''. 

In this section, we will focus on non-Hermitian Hamiltonians that represent balanced, spatially separated loss and gain. In a waveguide-array realization of such a Hamiltonian, one of the waveguides is  lossy, its parity-symmetric counterpart has gain, and the rest of the waveguides are neutral~\cite{Guo2009,Ruter2010}. To get a feel for properties of such a system and to define the terminology, let us start with the simplest example with $N=2$ waveguides. The tunneling Hamiltonian for this system is given by $H_t=-\hbar C (a^\dagger_1 a_2 + a^\dagger_2 a_1)$.  The non-Hermitian, $\mathcal{PT}$-symmetric potential, which represents gain in the first waveguide and loss in the second, is given by $V=i\hbar\gamma(a^\dagger_1 a_1 - a^\dagger_2 a_2)$. In a matrix notation, the total Hamiltonian becomes
\begin{equation}
\label{eq:h2by2}
H= \hbar\left[\begin{array}{cc}
i\gamma & - C \\
-C & -i\gamma \\
\end{array}\right] \neq H^\dagger.
\end{equation}
Although $H=H_t+V$ is not Hermitian, it is invariant under the combined parity ($\mathcal{P}: 1\leftrightarrow 2$) and time-reversal ($\mathcal{T}: i\rightarrow -i$) operations~\cite{qm2}. It is straightforward to obtain the eigenvalues $\lambda_\pm$ and (right) eigenvectors $|\pm\rangle_R$ of the Hamiltonian (\ref{eq:h2by2}). We remind the reader that since the matrix $H$ is not Hermitian, its left-eigenvectors and right-eigenvectors are not Hermitian conjugates of each other~\cite{ptajp,matrix}. 

For a small non-Hermiticity, $\gamma\leq C$, the eigenvalues of $H$ are purely real, and given by $\lambda_\pm=\pm\epsilon=\pm \hbar\sqrt{C^2-\gamma^2}$. The corresponding right-eigenvectors are given by 
\begin{equation}
\label{eq:evreal}
|\pm\rangle_R =\frac{1}{2}\left[|1\rangle\mp e^{\mp i\theta} |2\rangle\right],
\end{equation}
where $\sin\theta=\gamma/C\leq 1$. Thus, $_R\langle +|-\rangle_R =(-i)e^{i\theta}\sin\theta\neq 0$. Since the matrix $H$ is symmetric, $H=H^T$, the left-eigenvectors are obtained by taking the transpose of the right-eigenvectors. $|\pm\rangle_R$ are simultaneous eigenvectors of the combined $\mathcal{PT}$ operation as well, and each of them has equal weight on the gain and the loss site. When $\gamma=0$ the inner product is zero, whereas for $\gamma\rightarrow C$, the two eigenvalues become degenerate and {\it the two eigenvectors become parallel to each other}. For $\gamma\geq C$, the eigenvalues are purely imaginary complex conjugates, $\lambda_\pm=\pm i\hbar\Gamma=\pm i\hbar\sqrt{\gamma^2-C^2}$. The corresponding right-eigenvectors are now given by 
\begin{equation}
\label{eq:evimag}
|\pm\rangle_R=\frac{1}{\sqrt{1+ e^{\mp 2\phi}}} \left[|1\rangle +i e^{\mp\phi} |2\rangle\right],
\end{equation}
where $\cosh\phi=\gamma/C\geq 1$. Thus, the inner product of the two eigenvectors is equal to $1/\cosh\phi\leq 1$.  Note that now the eigenvectors are not simultaneous eigenvectors of the $\mathcal{PT}$-operation; the $|-\rangle_R$ eigenvector has higher weight on the gain site and the $|+\rangle_R$ eigenvector has higher weight on the loss site. 

The region of parameter space where all eigenvalues are real and the eigenvectors are simultaneous eigenvectors of the $\mathcal{PT}$ operation, $\gamma/C\leq 1$, is traditionally called the $\mathcal{PT}$-symmetric region, and $\gamma_{PT}=C$ is called the threshold loss-and-gain strength. For $\gamma/\gamma_{PT}>1$, complex conjugate eigenvalues emerge and the $\mathcal{PT}$-symmetry of the Hamiltonian $H$ is not shared by its eigenvectors with complex eigenvalues. Therefore, the emergence of complex eigenvalues is called $\mathcal{PT}$-symmetry breaking. In the following subsections, we present the properties of $N$-waveguide arrays with Hermitian, position-dependent tunneling profiles $C_\alpha(j)$ and a single pair of non-Hermitian, $\mathcal{PT}$-symmetric loss and gain potentials.  

%**************************************************************************************

\subsection{$\mathcal{PT}$ symmetric phase diagram}
\label{ss:ptphase}

% Typical PT-phase diagram for alpha-dependent lattice.
\begin{figure*}[htpb]
\begin{center}
\includegraphics[width=0.95\columnwidth]{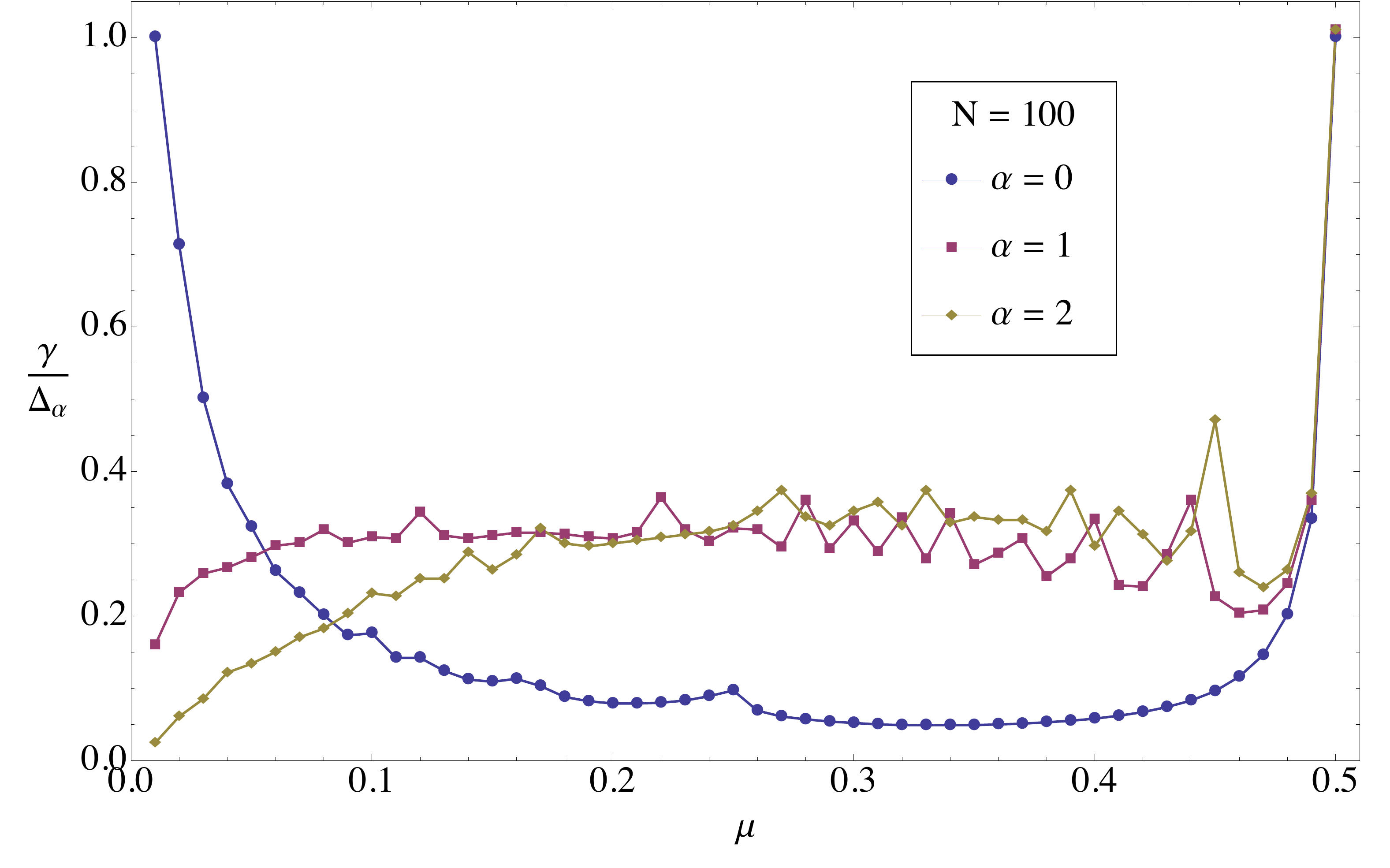}
\includegraphics[width=0.95\columnwidth]{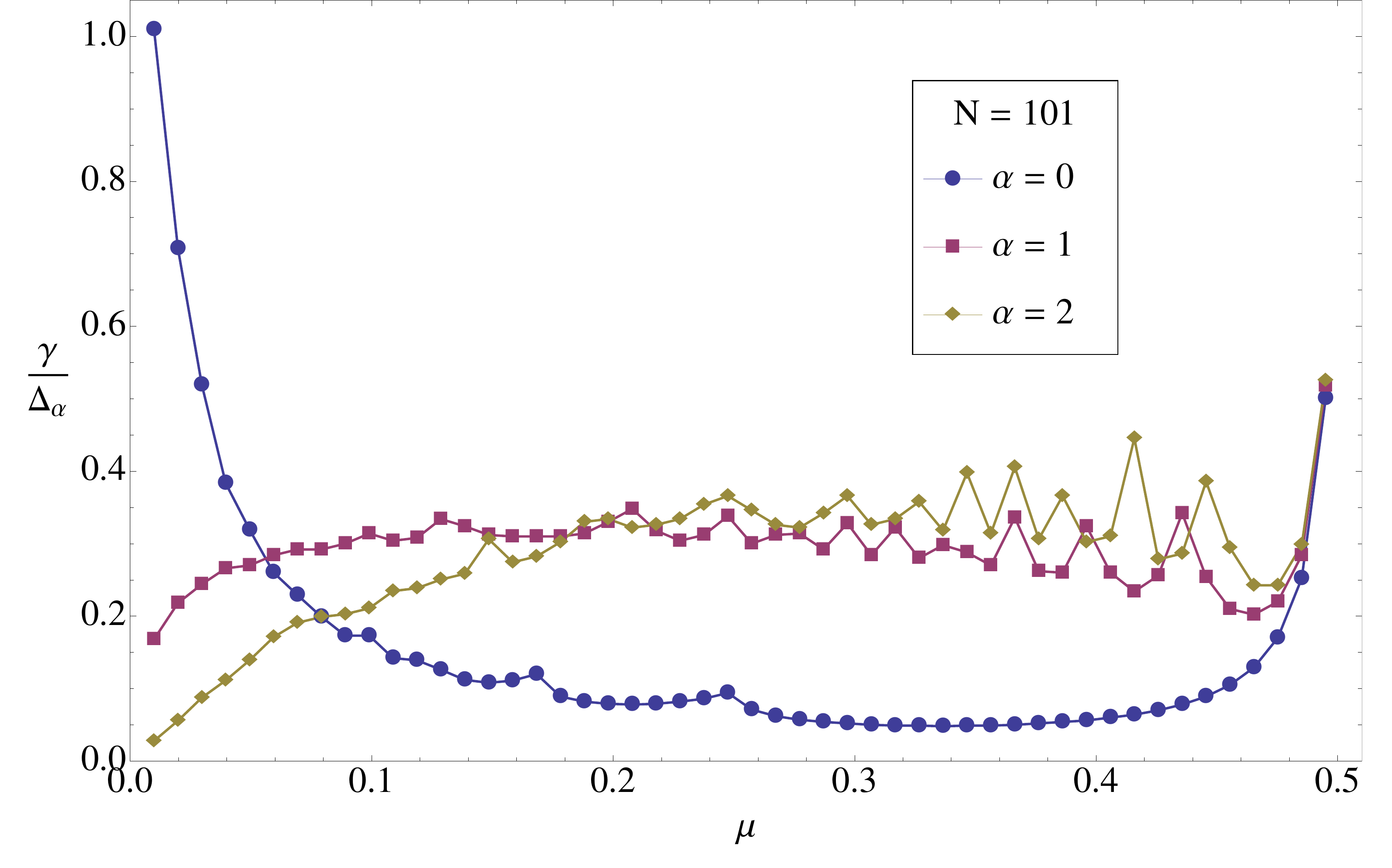}
\caption{$\mathcal{PT}$-symmetric phase diagram as a function of tunneling profile $\alpha\geq 0$. The vertical axes denote the strength of the non-Hermitian, loss (gain) term in units of the quarter-bandwidth $\Delta'_\alpha=\Delta_\alpha/(4\hbar)$ of the Hermitian lattice; the horizontal axes denote the relative position of the gain waveguide $\mu=m/N$. The left-hand panel, with even $N=100$, shows that $\gamma_{PT}(\mu)/\Delta'_\alpha=1$ is maximum at $\mu=0.5$, and remains relatively constant over a wide range of $\mu$ as $\mu\rightarrow 0$ for $\alpha\geq 1$. In contrast, for constant tunneling, the threshold loss-and-gain strength decays rapidly with decreasing $\mu$, but increases again as $\mu\rightarrow 0$. The right-hand panel shows corresponding, qualitatively similar, results for an odd array with $N=101$. For an odd array, the smallest separation between loss and gain waveguides is $D=2$, instead of $D=1$ in an even array, and therefore, the maximum threshold value for an odd array near $\mu=0.5$ is $\gamma_{PT}=0.5\Delta'_\alpha$.}
\label{fig:ptphase}
\end{center}
\vspace{-5mm}
\end{figure*}

We begin with the Hamiltonian for an $N$-site array with open boundary conditions, 
\begin{equation}
\label{eq:hpt}
H_{PT\alpha}=H_\alpha + i\gamma\left(a^\dagger_m a_m - a^\dagger_{\bar{m}} a_{\bar{m}}\right)
\end{equation} 
where $H_\alpha$ is the Hermitian tunneling Hamiltonian, Eq.(\ref{eq:halpha}), $1\leq m\leq N/2$ is the position of the waveguide with gain, and $\bar{m}=N+1-m$ is the parity-symmetric position of the waveguide with absorption. The parity operator in an array with open boundary conditions is given by $\mathcal{P}: a_k\rightarrow a_{\bar{k}}$. Thus, it follows that the Hermitian part of the Hamiltonian is $\mathcal{PT}$-symmetric, $C_\alpha(k)=C_\alpha(N-k)$, and so is the non-Hermitian potential term. Thus, to obtain the $\mathcal{PT}$-symmetric phase diagram, we need to obtain the eigenvalues of the Hamiltonian $H_{PT}$ and then locate the threshold loss and gain strength $\gamma_{PT}(\mu)$ as a function of the relative location $\mu=m/N$ of the gain waveguide. It is possible to obtain this threshold analytically only in the case of constant tunneling, $\alpha=0$~\cite{song,mark}; however, for an arbitrary $\alpha$, a numerical approach is most fruitful. By numerically tracking the emergence of complex eigenvalues of the tridiagonal matrix $H_{PT\alpha}$, we obtain the typical phase diagram, shown in Fig.~\ref{fig:ptphase}. Note that $\mu=1/N$ corresponds to largest distance between the loss and gain waveguides, whereas $\mu\sim 0.5$ corresponds to the shortest separation between them. Due to the constraint of parity-symmetric locations, in an even $N$-array this separation is unity, and for an odd $N$-array, the loss and gain locations have to be separated by a single waveguide between them. 

The left-hand panel in Fig.~\ref{fig:ptphase} shows the threshold strength measured in units of the 
lattice bandwidth as a function of relative location of the gain waveguide for an $N=100$ array with $\alpha=0$ (blue circles), $\alpha=1$ (red squares), and $\alpha=2$ (beige diamonds); all eigenvalues of $H_{PT\alpha}$ are real for values of $\gamma$ below the curve for that $\alpha$.  Note that we use quarter-bandwidth, $\Delta'_\alpha=\Delta_\alpha/(4\hbar)$, as the relevant scale in the phase diagram.  For $\alpha\geq 1$, the threshold strength is maximum $\gamma_{PT}/\Delta'_\alpha=1$ is at $\mu=0.5$, when the loss and gain waveguides are nearest neighbors. It reduces to $\gamma_{PT}/\Delta'_\alpha\sim 0.3$ and remains approximately constant for $0.15\leq \mu\leq 0.45$, and is monotonically suppressed with the separation $D=1+N(1-2\mu)$ between the loss and gain waveguides. Note that the behavior $\gamma_{PT}(\mu)$ for an array with constant tunneling amplitude, $\alpha=0$ is dramatically different. Starting from the maximum value of $\gamma_{PT}/C=1$ for closest loss and gain, the threshold strength first drops rapidly with increasing $d$, but is again enhanced as the loss and gain sites approach the two edges of the array. Thus, for moderate separations $\mu\sim 0.25$ and number of waveguides $N\sim 100$, the $\mathcal{PT}$-symmetric phase in an array with non-constant tunneling amplitudes is substantially stronger than in an array with constant tunneling amplitude. The right-hand panel shows the $\mathcal{PT}$-phase diagram for an array with an odd number of waveguides, $N=101$. We see that the robust nature of the $\mathcal{PT}$-symmetric phase for $\alpha\geq 1$ is maintained, although the threshold for smallest separation $\mu=(N-1)/2N$ is reduced to $\gamma_{PT}=0.5\Delta'_\alpha$~\cite{mark,derek}. 

We emphasize that although the qualitative form of the $\mathcal{PT}$-phase diagram is the same for different $N$, as $N$ increases, the threshold strength $\gamma_{PT}(\mu)/\Delta'_\alpha$ decreases for all separations except when the loss and gain are the closest ($\mu\sim 0.5$) or the farthest ($\mu=1/N$). Thus, rigorously, $\gamma_{PT}/\Delta'_\alpha(N)\rightarrow 0$ as $N\rightarrow\infty$; however, this is of no concern for experiments where the number of sites in an array - whether the ``site'' be an optical waveguide~\cite{Guo2009,Ruter2010,ptsynthetic}, an RLC circuit~\cite{ptelec}, or a pendulum~\cite{ptpendulum} - is typically $N\lesssim 100$. Since the $\mathcal{PT}$-symmetry breaking occurs when two adjacent eigenvalues $E_n,E_{n+1}$ become degenerate and then complex, and since the eigenvalues of $H_\alpha$ occur in pairs $\pm E_n$, it follows that, for a generic position $\mu$ of the gain waveguide, $N-4$ eigenvalues of the Hamiltonian $H_{PT\alpha}$ remain real while four eigenvalues become complex conjugate pairs. The remarkable exception to this rule is the case of nearest-neighbor loss and gain waveguides in an even $N$ array. In this case, since the array can be effectively divided into two systems, one with the loss and the other with the gain, all $N$ eigenvalues of $H_{PT\alpha}$ become complex simultaneously~\cite{derek,jake}. Thus, the implications of $\mathcal{PT}$-symmetry breaking are determined by both the threshold loss-and-gain strength $\gamma_{PT}$ and the location and number of eigenvalues that become complex at the threshold. 

%**************************************************************************************

\subsection{Time evolution across the $\mathcal{PT}$ threshold}
\label{ss:intensity}

% Time-dependent intensity evolution with gamma.
\begin{figure*}
\begin{center}
\includegraphics[width=\columnwidth]{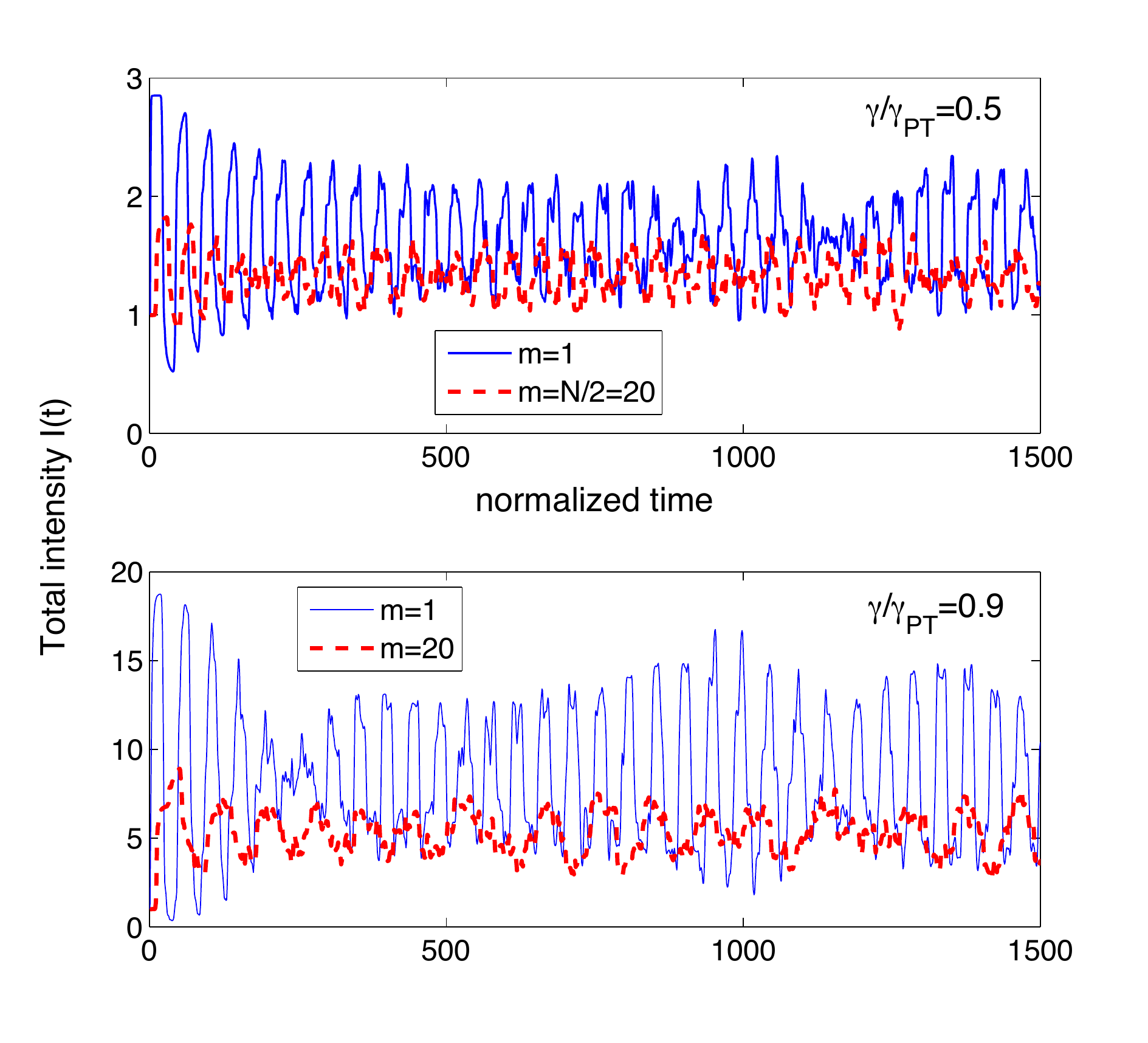}
\includegraphics[width=\columnwidth]{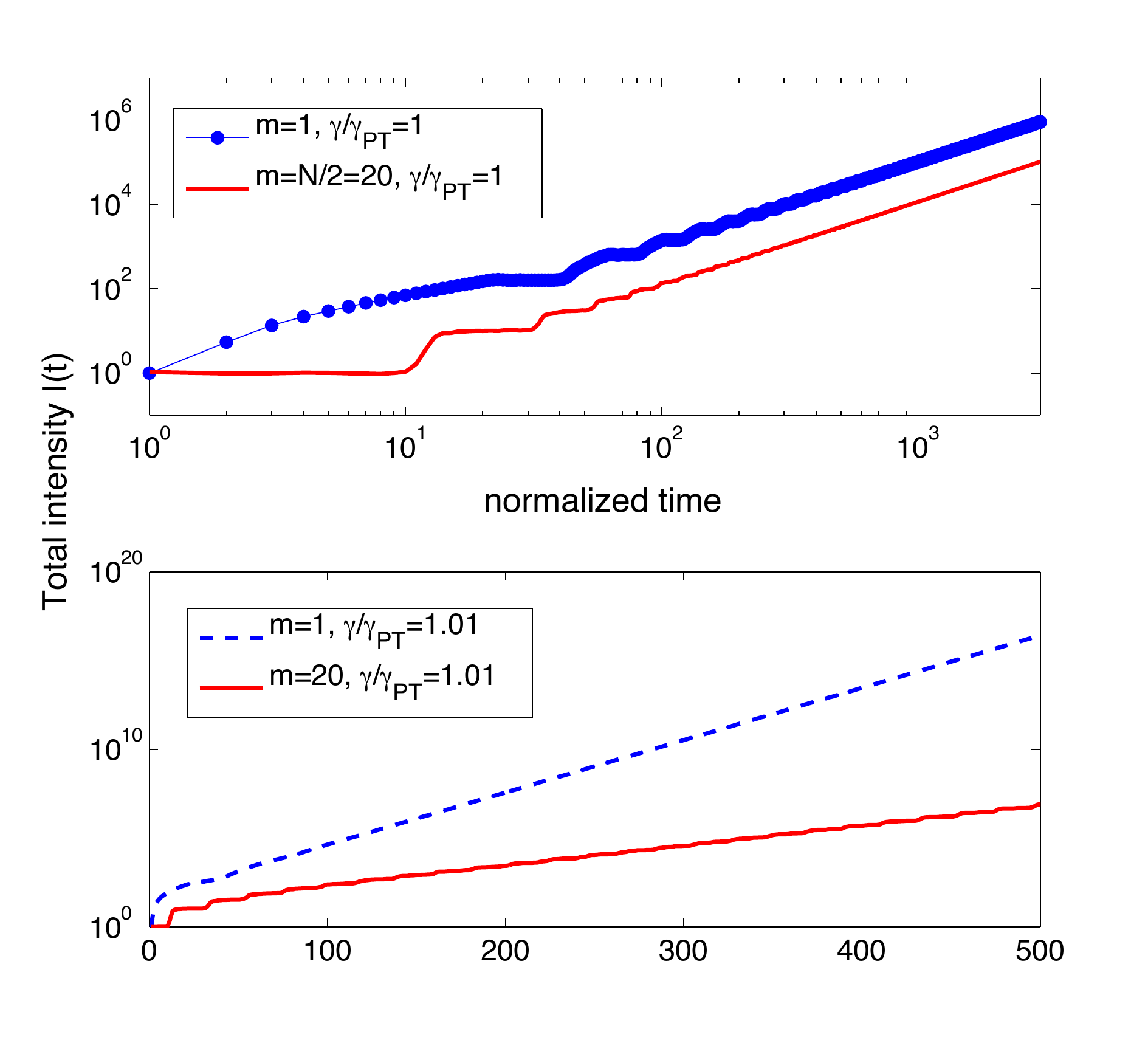}
\caption{Evolution of the time-dependent net intensity $I(t)$ as a function of loss-and-gain strength $\gamma/\gamma_{PT}$ in an $N=40$ waveguide array with the gain-waveguide at $m=1$ (blue curves) and $m=N/2$ (red curves). The left-hand panels show that $I(t)$ is remains bounded in the $\mathcal{PT}$-symmetric region, $\gamma/\gamma_{PT}<1$. The top-right panel shows that $I(t)\propto t^2$ at the threshold, $\gamma/\gamma_{PT}=1$. The bottom-right panel shows that in the $\mathcal{PT}$-symmetry broken region, $\gamma/\gamma_{PT}>1$, the net intensity diverges exponentially with time. These results, obtained for $|\psi(0)\rangle=|1\rangle$, have the same qualitative behavior for an arbitrary initial state.}
\label{fig:intensity}
\end{center}
\vspace{-5mm}
\end{figure*}
In the previous section with Hermitian Hamiltonians, we presented intensity profiles $I(p,t)$ for various initially normalized states, $\langle \psi(0)|\psi(0)\rangle=1$. Since the time evolution operator in these cases is unitary, $G^\dagger(t)=\exp\left[+i H^\dagger t/\hbar\right]=\exp\left[+i H t/\hbar\right] =G^{-1}(t)$, the total intensity of the time-evolved wave packet remains unity, $\sum_{p=1}^N I(p,t)=1$. For a non-Hermitian Hamiltonian, since $H_{PT}^\dagger\neq H_{PT}$, the corresponding time evolution operator is not unitary. Therefore, the norm of an initially normalized state is not preserved and the total intensity is a function of time, $I(t)=\sum_{p=1}^N I(p,t)\neq 1$. Note that $G(t)=\exp\left[-iH_{PT}t/\hbar\right]$ is not a unitary operator irrespective of whether the system is in the $\mathcal{PT}$-symmetric phase or has complex conjugate eigenvalues. 

To get a better feel for this non-unitary time evolution operator, let us calculate it for the two-site Hamiltonian, Eq.(\ref{eq:h2by2}). From the completeness property of its left and right eigenvectors, it follows that
\begin{equation}
\label{eq:gpt}
G(t) = |+\rangle_R e^{-i \lambda_{+}t/\hbar} {_L\langle +|}  + |-\rangle_R e^{-i\lambda_{-} t/\hbar} {_L\langle -|} 
\end{equation}
where the left eigenvectors $_L\langle\pm |$ are obtained by transposing the right eigenvectors $|\pm\rangle_R$. In the $\mathcal{PT}$-symmetric phase, $\gamma/C\leq 1$, Eq.(\ref{eq:evreal}) implies that 
\begin{equation}
\label{eq:gless}
G_{\leq}(t)=\left[\begin{array}{cc}
\cos\tau+\frac{\gamma}{\epsilon}\sin\tau & +i\frac{C}{\epsilon}\sin\tau \\
+i\frac{C}{\epsilon}\sin\tau & \cos\tau-\frac{\gamma}{\epsilon}\sin\tau\\
\end{array}\right]=G^T_\leq(t)
\end{equation}
where $\tau=\epsilon t/\hbar$ is the dimensionless time. We leave it to the reader to verify that $G_\leq(t)$ is not unitary, but its eigenvalues have unit modulus and are given by $e^{\pm i\tau}$. 
Therefore the non-unitary time evolution operator satisfies $\det G_\leq(t)=1$. In the $\mathcal{PT}$-symmetry broken phase, $\gamma/C\geq 1$, a corresponding calculation using Eq.(\ref{eq:evimag}) gives
\begin{equation}
\label{eq:ggreater}
G_\geq(t)=\left[\begin{array}{cc}
\cosh\tau' +\frac{\gamma}{\Gamma}\sinh\tau' & i\frac{C}{\Gamma}\sinh\tau'\\
i\frac{C}{\Gamma}\sinh\tau' &  \cosh\tau' -\frac{\gamma}{\Gamma}\sinh\tau'\\
\end{array}\right]
\end{equation}
where $\tau'=\Gamma t/\hbar$. The reader can verify that $G_\geq(t)$ is not unitary, its eigenvalues are $e^{\pm\tau'}$, and thus, $\det G_\geq(t)=1$. 

We note that the matrix elements of $G_\leq(t)$ are bounded, those of $G_\geq(t)$ diverge with increasing time, and that the time evolution operator is continuous across the $\mathcal{PT}$-symmetry threshold. At the threshold $\gamma=C$, since the Hamiltonian is singular, $H^2=0$, the exponential expansion for the time-evolution operator truncates at the linear order and gives
\begin{equation}
\label{eq:gth}
G_{C}(t)=\left[\begin{array}{cc} 
1+ Ct/\hbar & i Ct/\hbar\\
i Ct/\hbar & 1- Ct/\hbar\\
\end{array}\right].
\end{equation}
Since the time evolved state is given by $|\psi(t)\rangle=G(t)|\psi(0)\rangle$, the change in net intensity is proportional to unitary deficit, $G^\dagger(t) G(t)-1$. Equations (\ref{eq:gless})-(\ref{eq:gth}) show that, for a $\mathcal{PT}$-symmetric Hamiltonian (\ref{eq:h2by2}), the net intensity $I(t)$ in the $\mathcal{PT}$-symmetric phase remains bounded, increases exponentially with time in the $\mathcal{PT}$-symmetry broken phase, and exactly at the threshold,  varies quadratically with time at long times~\cite{zheng2010}. 

Figure~\ref{fig:intensity} shows the evolution of net intensity $I(t)$ in an $N=40$ waveguide array with constant tunneling, $\alpha=0$, the loss-and-gain waveguides farthest apart ($m=1$) or closest together ($m=N/2=20$) as a function of $\gamma/\gamma_{PT}$. These numerically obtained results are for an initial state localized at the first waveguide, $|\psi(0)\rangle=|1\rangle$. We remind the reader that the crucial difference between the $m=1$ case and the $m=N/2$ case is that only four eigenvalues, at the center of the cosine-band become complex for $m=1$, whereas all eigenvalues simultaneously become complex when $m=N/2$~\cite{mark,derek,jake}. The top-left and bottom-left panels show that in the $\mathcal{PT}$-symmetric phase, $\gamma/\gamma_{PT}<1$, the net intensity $I(t)$ oscillates but remains bounded, and its time-average increases monotonically with its proximity to the $\mathcal{PT}$-symmetric phase boundary. In addition, they show that the average and fluctuations in the $m=N/2$ case are smaller than those in the $m=1$ case. The top-right panel shows $I(t)$ at the threshold, $\gamma/\gamma_{PT}=1$, for the two cases; note the logarithmic scale on both axes. At small times, the order-of-magnitude difference between intensities for the two gain-waveguide locations is consistent results in the left-hand panels. At longer times, we see that the net intensity scales quadratically with time, although the prefactor of this quadratic dependence is greater for the $m=1$ case. The bottom-right panel shows $I(t)$ in the $\mathcal{PT}$-symmetry broken phase, $\gamma/\gamma_{PT}=1.01$; note the logarithmic scale on the vertical axis. These results show that, as expected, the net intensity diverges exponentially, but with a larger exponent for loss and gain waveguides at the two ends of the array, $m=1$. We emphasize that this qualitative trend is valid for arbitrary location and shape of the initial wave packet. The results in Fig.~\ref{fig:intensity} show that the simple $2\times 2$ non-Hermitian Hamiltonian, Eq.(\ref{eq:h2by2}), captures the time-dependence of the net intensity in a large tight-binding array, although it does not capture the full gamut of $\mathcal{PT}$-symmetry breaking signatures~\cite{derek}. 

%**************************************************************************************

\subsection{Intensity correlations with Hermitian or $\mathcal{PT}$-symmetric disorders}
\label{ss:ptdisorder}
We have seen in Sec.~\ref{ss:disorder} that (Hermitian) disorder leads to ``localization'' of an arbitrary initial state, that is characterized by a steady-state, disorder-averaged intensity profile $\langle I(p)\rangle$. The steady-state intensity profile is solely determined by the initial state and the strength of the disorder potential, but is independent of whether the disorder is in the on-site potentials or tunneling amplitudes. Therefore, the site-dependent steady-state intensity measurements can only determine the strength of the disorder, but not the type of the disorder. These two disorders affect the particle-hole symmetric spectrum of the clean lattice in qualitatively different manners: the on-site, diagonal disorder destroys this symmetry whereas the tunneling, off-diagonal disorder preserves it. Therefore, although intensity measurements are insensitive to it, it is known that intensity correlation function is able to distinguish between the on-site and tunneling disorders~\cite{hbtcorr}.  

% Intensity correlations in PT/off-diagonal disorder. 
\begin{figure}[htbp]
\begin{center}
\includegraphics[width=\columnwidth]{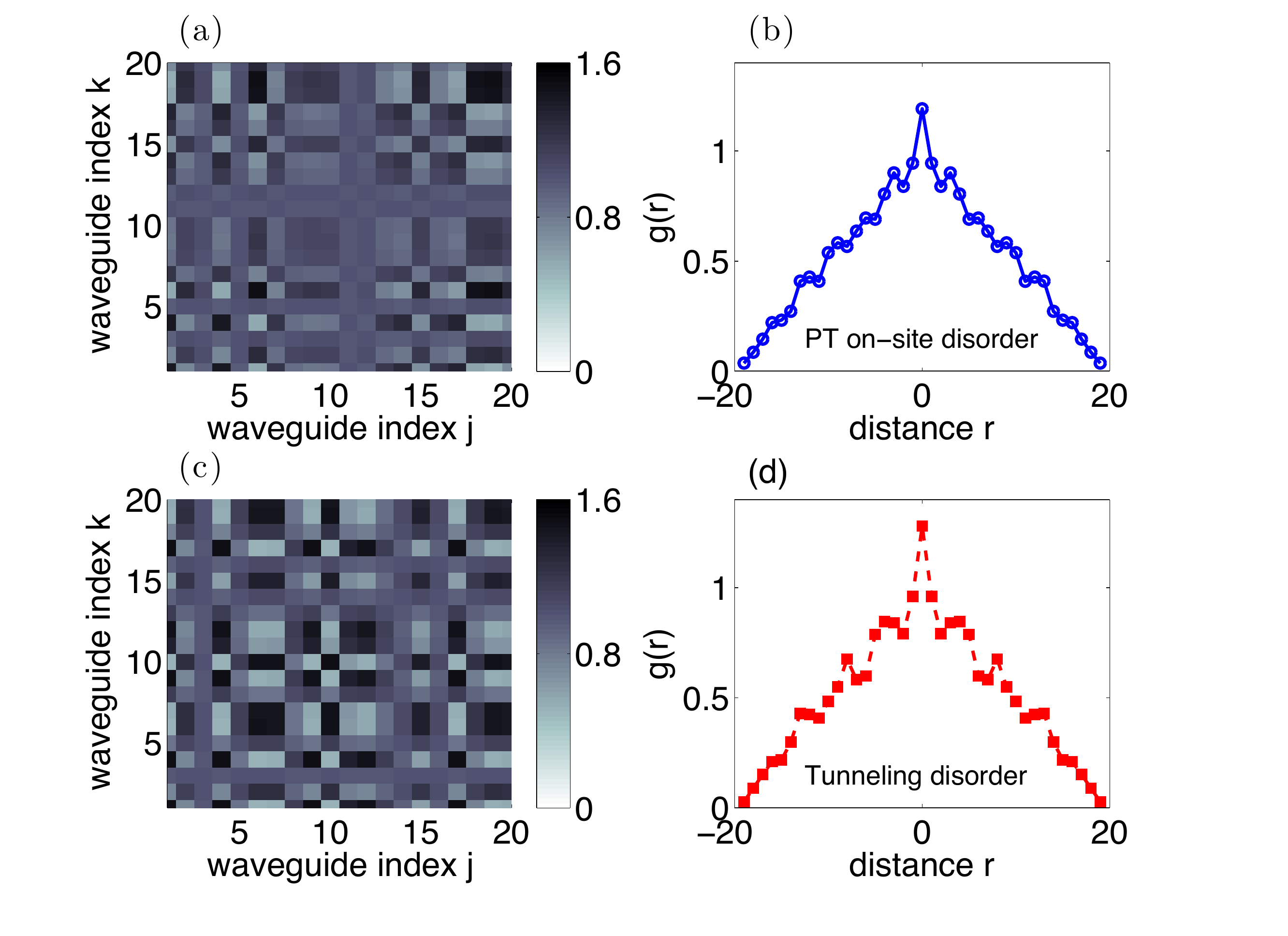}
\caption{Normalized correlation matrix $\Gamma_{jk}$ and intensity correlation function $g(r)$ for an $N=20$ array with constant tunneling, $\mathcal{PT}$-symmetric on-site disorder (top line) and Hermitian, tunneling disorder (bottom line) with zero mean and equal variance $v_d=0.02\Delta_B$. The steady-state $\Gamma_{jk}$, panels (a) and (c), are different for the two sources of disorder, whereas the steady-state intensity correlation function $g(r)$, panels (b) and (d), is insensitive to them. Their similarity shows that the particle-hole symmetry of the disordered spectrum is instrumental to the correlation function properties.}
\label{fig:hbtdisorder}
\end{center}
\vspace{-5mm}
\end{figure}
In contrast to the Hermitian potential, a non-Hermitian, $\mathcal{PT}$-symmetric potential, 
in the $\mathcal{PT}$-symmetric phase, preserves particle-hole symmetry of the resulting, purely real spectrum~\cite{Joglekar2010a}. Therefore, in this section, we compare the steady-state intensity correlations from a Hermitian tunneling disorder and a non-Hermitian, $\mathcal{PT}$-symmetric disorder, both with zero mean and equal variance. The $\mathcal{PT}$-symmetric disorder potential is given by
\begin{equation}
\label{eq:vpt}
V=\sum_{m=1}^{N/2} i\gamma_m\left(a^\dagger_m a_m - a^\dagger_{\bar{m}} a_{\bar{m}}\right)
\end{equation}
where the random, loss (or gain) potentials $|\gamma_m|\leq\gamma_{PT}(\mu=m/N)$ ensure that the system is in the $\mathcal{PT}$-symmetric phase. The normalized correlation matrix is defined as~\cite{hbtcorr}
\begin{equation}
\label{eq:hbt}
\Gamma_{jk}(t)=\frac{\langle I(j,t) I(k,t)\rangle}{\langle I(j,t)\rangle\langle I(k,t)\rangle}\bigg\vert_{t\gg 1}
\end{equation}
where $I(j,t)$ is the intensity profile determined by the initial state $|\psi(0)\rangle$ and the disorder potential. $\langle I(j,t)\rangle$ is the disorder-averaged intensity that becomes independent of time at long times (Sec.~\ref{ss:disorder}). The net intensity $\sum_p I(p,t)$ is conserved at unity for a Hermitian disorder, but not for the $\mathcal{PT}$-symmetric disorder. The intensity correlation function is defined as 
\begin{equation}
\label{eq:g2} 
g(r)=\frac{1}{N}\sum_{j=1}^N\Gamma_{j,j+r},
\end{equation}
and represents the sum of weights along a diagonal that is shifted by $r$ from the main diagonal of the steady-state correlation matrix, Eq.(\ref{eq:hbt}). Figure~\ref{fig:hbtdisorder} shows the normalized, steady-state correlation matrix $\Gamma_{ij}$ and the intensity correlation function $g(r)$ for an $N=20$ array with constant tunneling, $C_\alpha(j)=C$, and initial state $|\psi(0)\rangle=(|9\rangle+|10\rangle)/\sqrt{2}$. The top line shows the results for a $\mathcal{PT}$-symmetric, on-site disorder, whereas the bottom line has results for a Hermitian, tunneling disorder; both disorders have zero mean, equal variance $v_d/\Delta_B=0.02$, and the results are averaged over $M\sim 10^4$ disorder realizations. Panels (a) and (c) show that the full correlation matrix $\Gamma_{jk}$ is sensitive to the source of disorder. However, panels (b) and (d) show that the intensity correlation function $g(r)=g(-r)$ cannot distinguish between the two. Thus, symmetry properties of the disorder-induced spectrum are reflected in the disorder-averaged intensity correlation function, and not the on-site or off-diagonal nature of disorder~\cite{clintdisorder}. 

These results also suggest that although intensity distribution, or intensity correlation function is insensitive to the disorder distribution function, higher order intensity correlations may encode signatures of different disorder distributions that have zero mean and identical variance~\cite{Thompson2010,karr2011}. 

%**************************************************************************************
%**************************************************************************************

\section{Conclusion}
\label{sec:disc}
In this article, we have presented the properties of coupled waveguide arrays. We have argued that they provide a versatile and robust realization of a tight-binding model, ideally suited for investigating many quantum, quasi-classical, and bandwidth effects that are not easily accessible in ``naturally occurring'' lattices in electronic materials. We have shown that finite arrays with small number of waveguides exhibit a rich variety of effects, such as localization in the parity-symmetric waveguide, that are absent in a lattice with sites $N\rightarrow\infty$. 

Due to the ease of introducing absorption or amplification, coupled optical waveguides are also well-suited to model open systems with spatially separated, balanced loss and gain. Such systems are formally described by non-Hermitian, $\mathcal{PT}$-symmetric Hamiltonian. Since the spectrum of such Hamiltonian changes from purely real to complex, and since the time-evolution under such Hamiltonian is always non-unitary, we have discussed a few salient properties of $\mathcal{PT}$-symmetric lattice models. 

In this review, we have ignored nonlinear effects that arise at high intensities in a waveguide, and that are expected to play a large role in the $\mathcal{PT}$-symmetry broken region where the net intensity increases exponentially with time. We have not considered the effects of shape-preserving solitonic solution that exist in the nonlinear regime on time evolutions discussed here. In addition, we have not discussed the effects of $\mathcal{PT}$-symmetric, non-Hermitian disorder, including the fate of Anderson localization, in the $\mathcal{PT}$-symmetry broken region. The investigation of these outstanding questions will further deepen our knowledge of this exciting research area. 

%**************************************************************************************
%**************************************************************************************

\section*{Acknowledgments}
This work was supported by the NSF DMR-1054020 (Y.J.), and a GAANN Fellowship (C.T.) from the US Department of Education grant (G.V.)

%**************************************************************************************
%**************************************************************************************

%\clearpage

% BibTeX users please use
\bibliographystyle{epj}
%\bibliography{Sources}

%**************************************************************************************
%**************************************************************************************

\end{document}